\documentstyle[12pt,a4,fleqn]{article}
\makeatletter
\newdimen\@bls
\@bls=\baselineskip

\def\section{\setcounter{equation}{0}
  \@startsection{section}{1}{\z@}{1.5\@bls
  plus .4\@bls minus .1\@bls}{\@bls}{\large\bf}}
\def\subsection{\@startsection{subsection}{2}{\z@}{\@bls
  plus .3\@bls minus .1\@bls}{\@bls}{\large\sl}}
\def\subsubsection{\@startsection{subsubsection}{3}{\z@}{\@bls
  plus .2\@bls}{0.0001pt}{\normalsize\it}}
\def\paragraph{\@startsection{paragraph}{4}{\z@}{3.25ex plus
  2ex minus 0.2ex}{-1em}{\normalsize\bf}}
\makeatother
\def\mod{\mathop{\rm mod}}
\def\Tr{\mathop{\rm Tr}}
\def\sgn{\mathop{\rm sgn}}

\let\e=\varepsilon

\begin{document}
%%%%%%%%%%%%%%%%%%%%%%%%%%%%%%%%%%%%%%%%%%%%%%%%%%%%%%%%%%%%%%%%%%%%%%%
\begin{titlepage}
\rightline{PDMI PREPRINT --- 19/1997}
\vspace*{1in}
\begin{flushleft}
{\Large\bf
Temperature correlators in the\\[8pt]
two-component one-dimensional gas
}\\[0.5in]
{\sc  A.\,G.\,Izergin\footnote{\rm E-mail: izergin@pdmi.ras.ru},
A.\,G.\,Pronko\footnote{\rm E-mail: agp@pdmi.ras.ru}}\\[.5in]
{\it Sankt-Petersburg Department
of V.\,A.\,Steklov Mathematical Institute,\\
Fontanka 27, 191\,011 Sankt-Petersburg, Russia}.
\end{flushleft}
\vspace{1in}
\hrule
\vspace{.2in}
\noindent
{\bf Abstract}
\bigskip

The quantum nonrelativistic
two-component Bose and Fermi gases with the infinitely strong
point-like coupling between particles in one space dimension are
considered. Time and temperature dependent correlation functions
are represented in the thermodynamic limit as Fredholm determinants of
integrable linear integral operators.
\vspace{.2in}
\hrule
\vfill
\end{titlepage}
%%%%%%%%%%%%%%%%%%%%%%%%%%%%%%%%%%%%%%%%%%%%%%%%%%%%%%%%%%%%%%%%%%%%%%%
\section*{Introduction}

The recent progress in calculating correlation functions of quantum
solvable models is based on the fact that they are governed by
classical integrable differential equations. That the language of
classical differential equations is quite natural for the description
of quantum correlation functions was realized a time ago
\cite{Tr1,Tr2,Tr3,Jimbo}.  The idea of the approach suggested in
\cite{IIK-89,IIKS-90,IIKS-93} is to consider the Fredholm determinant
in the representation for a correlation function of a quantum
integrable model as a tau-function for a classical integrable system
(see also the book \cite{KBI} where the results for the simplest model
of one-dimensional impenetrable bosons are reviewed). The necessary
first step (which is also of interest by itself) in this approach is to
represent the correlation function as the Fredholm determinant of a
linear integral operator of a special kind (the ``integrable'' integral
operator).  It is to be mentioned that the first determinant
representation of this kind was given in \cite{L-64,L-66} for the
equal-time temperature correlators of the impenetrable bosons in one
space dimension.

We consider the exactly solvable one-dimensional model of the
nonrelativistic two-component Bose and Fermi gases with the infinitely
large coupling constant, $c=\infty$. This model for any values of $c$
was solved in the paper of Yang \cite{Yang} where the eigenvectors of
the Hamiltonian were constructed using relations known now as the
famous Yang-Baxter relations. The general approach to the solution of
the model with $n$ internal degrees of freedom was given in paper
\cite{Sutherland}, see also \cite{Gaudin} (for the solution in the
frame of the algebraic Bethe Ansatz see paper \cite{Kulish}).  The
Fredholm determinant representation for the equal-time temperature
correlators of the Fermi gas with $c=\infty$ was derived in paper
\cite{Berkovich} where also the long distance asymptotics was
constructed.

In our paper the Fredholm determinant representations for the time
dependent temperature correlation functions are obtained, for both Bose
and Fermi gases with $c=\infty$. To do this, we generalize the method
used earlier for the one-component models
\cite{KS-90,KS-91,CIKT-92,CIKT-93}.  A presentation of our results was
given in paper \cite{IP-97}.

The contents of this paper is as follows.  In Section 1, the explicit
expressions for eigenstates (the nested Bethe eigenvectors) are given.
At $c=\infty$ the auxiliary lattice problem for the nested Bethe Ansatz
is reduced to the problem of finding eigenvectors for the cyclic shift
operator and one can use the eigenvectors of the Heisenberg $XX0$ chain
to solve it. Due to the simple form of the $XX0$ basis, all further
calculations can be done explicitly.  The form factors of the model in
the finite box are calculated in Section 2.  The normalized mean values
of the products of two canonical fields with respect to Bethe
eigenstates are obtained in Section 3.  In Section 4, the expressions
for the two point temperature correlation functions in the
thermodynamic limit are derived which is the main result of our paper.
Important particular cases are discussed in Section 5.
%%%%%%%%%%%%%%%%%%%%%%%%%%%%%%%%%%%%%%%%%%%%%%%%%%%%%%%%%%%%%%%%%%%%%%%
\section{The eigenstates of the Hamiltonian}

The model in the finite box of length $L$ with the periodic boundary
conditions is defined by the secondary quantized Hamiltonian
\begin{equation} \label{H}
H=\int\limits_{0}^{L} dx \left\{
(\partial_x\psi^+ \partial_x \psi)
+c :(\psi^+ \psi)^2:
-h (\psi^+ \psi)
+B (\psi^+\sigma^z\psi) \right\},
\end{equation}
where $c$ is a coupling constant, $h$ is a chemical potential,
$B$ is a constant external transverse field and $:\ :$ means the normal
ordering. The Pauli matrices are normalized as
$[\sigma^x,\sigma^y]=2i\sigma^z$. The one-dimensional fields
$\psi_\alpha^{}(x)$ and $\psi_\alpha^+(x)$ ($\alpha = 1,2$) are quantum
operators in the Fock space with the canonical commutation relations
for the Bose gas ($\e =+1$ in the equations below) and with the
canonical anticommutation relations for the Fermi gas ($\e = -1$):
\begin{eqnarray} \label{ecom}
\lefteqn{
\psi^{}_\alpha(x) \psi^+_\beta(y)
-\e\,\psi^+_\beta(y) \psi^{}_\alpha(x)
=\delta_{\alpha\beta}\delta(x-y),
}
\nonumber\\
\lefteqn{
\psi^+_\alpha(x) \psi^+_\beta(y)
-\e\,\psi^+_\beta(y) \psi^+_\alpha(x)
=0,\quad
\psi^{}_\alpha(x) \psi^{}_\beta(y)
-\e\,\psi^{}_\beta(y) \psi^{}_\alpha(x)
=0.
}
\end{eqnarray}
In what follows we consider both cases simultaneously. It appears that the
dependence of the correlation functions on the statistics is  rather
simple.

The two-point temperature correlation functions are defined as the
temperature normalized mean values,
\begin{eqnarray} \label{T-cor}
\lefteqn{
G^{(-)}_\alpha(x,t;h,B)
=\frac{\Tr\left[e^{-H/T}
\psi^+_\alpha(x,t)\psi^{}_\alpha(0,0)\right]}
{\Tr\left[e^{-H/T}\right]},
}
\nonumber\\
\lefteqn{
G^{(+)}_\alpha(x,t;h,B)
=\frac{\Tr\left[e^{-H/T}
\psi^{}_\alpha(x,t)\psi^+_\alpha(0,0) \right]}{
\Tr\left[e^{-H/T}\right]},
}
\end{eqnarray}
where
\begin{equation} \label{Time}
\psi^+_\alpha (x,t)=e^{itH}\psi^+_\alpha(x) e^{-itH},\quad
\psi^{}_\alpha (x,t)=e^{itH}\psi^{}_\alpha(x) e^{-itH}
\end{equation}
and the traces are taken in the whole Fock space.
Correlators $G_1$ and $G_2$ are related by changing the sign of the
magnetic field,
\begin{equation} %\label{}
G^{(\pm)}_1(x,t;h,B)=G^{(\pm)}_2(x,t;h,-B),
\end{equation}
so that it is sufficient to calculate the correlators
$G^{(\pm)}_1(x,t;h,B)$ only. Our aim is to
calculate the correlators in the thermodynamic limit, i.e., at
$L\to\infty$ keeping $h$ and $B$ fixed. To do this, we perform first
the calculation for the finite box and then go to the limit.

The basis in the Fock space of the model is constructed by acting
with operators $\psi^+_\alpha(x)$ onto the pseudovacuum
$|0\rangle$ defined as
\begin{equation} %\label{}
\psi^{}_{\alpha}(x)|0\rangle=0, \quad
\langle 0|\psi^+_{\alpha}(x)=0, \quad
\langle 0|0\rangle=1.
\end{equation}
We say that a state belongs to the sector ($N,M$) of the Fock space if
it contains $N-M$ particles of type 1 ($\alpha = 1$) and $M$ particles
of type 2 ($\alpha = 2$), so that the total number of particles is $N$.
The number of particles of each type being conserved separately, an
eigenstate of the Hamiltonian can be obtained as a linear superposition
of the basis states from the same sector. In the sector ($N,M$) the
eigenstates are enumerated by two sets, $\{k\}=k_1,\ldots,k_N$ and
$\{\lambda\}=\lambda_1,\ldots,\lambda_M$, of unequal (in each set
separately) real numbers. Thus, the eigenstates in the sector $(N,M)$
can be written in the form
\begin{eqnarray} \label{PsiNM}
\lefteqn{
|\Psi_{N,M}(\{k\};\{\lambda\})\rangle
=\int\limits_{0}^{L} dz_1 \ldots \int\limits_{0}^{L} dz_N
\sum_{\alpha_1,\ldots,\alpha_N=1,2}^{}
\chi_{N,M}^{\alpha_1\ldots\alpha_N}(z_1,\ldots,z_N|\{k\};\{\lambda\})
}
\nonumber\\ &&\times
\psi^+_{\alpha_1}(z_1)\cdots \psi^+_{\alpha_N}(z_N) |0\rangle,
\end{eqnarray}
where the wave function
$\chi_{N,M}^{\alpha_1\ldots\alpha_N}(z_1,\ldots,z_N|\{k\};\{\lambda\})$
is not equal to zero only if $M$ elements are equal to 2 and
$N-M$ elements are equal to 1 in the set
$\alpha_1,\ldots,\alpha_N$, i.e.,  $\sum_{j=1}^{N} \alpha_j = N+M$.

The wave functions for the finite coupling constant $c$ were obtained
in \cite{Yang} (see also \cite{Gaudin}).
After imposing the periodic boundary conditions, the eigenfunctions of
the Hamiltonian are given by the two-component (nested) Bethe Ansatz.
The second component of the nested Ansatz (the ``auxiliary lattice
problem'') is the Bethe Ansatz for the inhomogeneous $XXX$ chain,
which gives the higher weight eigenvectors. The complete basis is then
obtained by acting with the Yangian generators onto the Bethe Ansatz
vectors.

The model at $c=\infty$ is defined by taking the limit $c\to\infty$ in
the expressions for the wave functions. This can be done differently.
One way to do this is to take the limit in the final expressions for
the Bethe eigenfunctions, keeping the $XXX$ chain eigenfunctions for
auxiliary lattice problem.  We use another way. It is to take the limit
in the expressions for the wave functions before imposing the periodic
boundary conditions, and to satisfy the periodic boundary conditions
after that.  On this way the auxiliary lattice problem is reduced to
the eigenstate problem for the cyclic shift operator on the lattice
which can be solved explicitly, e.g., using eigenfunctions of the $XX0$
chain with the periodic boundary conditions. The advantage of this
formulation is also the completeness of the $XX0$ eigenfunctions in the
isospin space.

The Yang's wave functions for $c=\infty$ can be written in the form
\begin{eqnarray} \label{chi}
\lefteqn{
\chi_{N,M}^{\alpha_1\ldots\alpha_N}(z_1,\ldots,z_N|\{k\};\{\lambda\})
}
\nonumber\\ &&
= {1 \over N!}
\left[\sum_{P}^{} (-\e)^{[P]}
\xi^{\alpha_{P_1}\ldots\alpha_{P_N}}_{N,M}(\{\lambda\})\
\theta(z_{P1}<\ldots<z_{PN}) \right]
{\det}_N\{e^{ik_a z_b}\},
\end{eqnarray}
where $\xi^{\alpha_1\ldots\alpha_N}_{N,M}(\{\lambda\})$ are
components of a $2^N$-dimensional vector $\xi_{N,M}(\{\lambda\})$.
The sum in (\ref{chi}) is taken over all the permutations of
$N$ numbers, $P:(1,2,\ldots,N) \to (P_1,P_2,\ldots,P_N)$; $[P]$ denotes
the parity of the permutation.

The periodic boundary conditions for the wave function
give the auxiliary lattice problem which is much simpler for
$c=\infty$ than in the general case:
\begin{equation} \label{C_N}
\xi^{\alpha_1\alpha_2\ldots\alpha_N}_{N,M}(\{\lambda\})
=(-\e)^{N+1}e^{ik_j L}
\xi^{\alpha_2\ldots\alpha_N\alpha_1}_{N,M}(\{\lambda\}).
\end{equation}
These equations should hold for all quasimomenta $k_1,\ldots,k_N$.
They can be regarded as the eigenvalue problem for the
cyclic shift operator $C_N$ acting in  a $2^N$-dimensional vector space
which can be identified with the space of states for the
spin-$\frac{1}{2}$ chain with $N$ sites.

The vectors $\xi_{N,M}$, which have to be eigenvectors of cyclic shift
operator $C_N$, can be chosen as the eigenvectors of the $XX0$
Heisenberg spin-$\frac{1}{2}$ chain with the periodic boundary
conditions.  We write the expressions for $\xi_{N,M}$ in the form used
in \cite{CIKT-93}:
\begin{eqnarray} \label{XX0}
\lefteqn{
\xi_{N,M}(\{\lambda\})
=\sum_{n_1=1}^{N}\ldots \sum_{n_M=1}^{N}
\varphi_{N,M}(n_1,\ldots,n_M|\{\lambda\})\
\sigma^{-}_{(n_1)}\cdots \sigma^{-}_{(n_M)}|\!\Uparrow_N),
}
\nonumber\\
\lefteqn{
\varphi_{N,M}(n_1,\ldots,n_M|\{\lambda\})
={1 \over M!}
\left[\prod_{1\le j<l \le N}^{} {\rm sgn}(n_l-n_j) \right]
{\det}_M \{e^{i\lambda_a n_b}\},
}
\end{eqnarray}
where $|\!\Uparrow_N)=\otimes_{n=1}^{N}|\!\uparrow_{n})$
(all spins up) is the pseudovacuum for the chain with $N$ sites,
$(\Uparrow_N\!|\!\Uparrow_N)=1$.
The eigenvalues of $C_N$ on the vectors (\ref{XX0})
are  $\ \exp\left(i \sum_{b=1}^{M}\lambda_b\right)$.

The periodic boundary conditions for the function $\varphi_{N,M}$
together with equation (\ref{C_N}) result in
the system of the Bethe equations,
\begin{equation} \label{Betheeqns}
\begin{array}{ll}
e^{ik_a L}=(-\e)^{N-1} e^{i\sum_{b=1}^{M}\lambda_b}, & a=1,\ldots,N,\\
e^{i\lambda_b N}=(-1)^{M+1},                         & b=1,\ldots,M.
\end{array}
\end{equation}
The permitted values
$(k_a)_j\in(-\infty,+\infty)$ of the quasimomenta
$k_a$ ($a=1,\ldots,N$) are
\begin{equation} \label{k_j}
(k_a)_j=\frac{2\pi}{L}\left(-\Bigl(\frac{1+\e}{2}\Bigr)
\frac{N-1}{2}+j\right)+
\frac{1}{L}\sum_{b=1}^{M}\lambda_b, \quad j \in Z,
\end{equation}
while the permitted values $(\lambda_b)_l \in (-\pi,\pi]$ of the
quasimomenta $\lambda_b$ ($b=1,\ldots,M$) are
\begin{equation} \label{lambda_l}
(\lambda_b)_l=
\frac{2\pi}{N}\left(-\frac{N}{2}-\frac{1+(-1)^{N-M}}{4}
+l\right),
\quad l=1,\ldots,N.
\end{equation}
Each quasimomentum $k_a$ (\ref{k_j}) is a function of the
total quasimomentum $\sum_{b=1}^{M}\lambda_b$ of the corresponding
auxiliary lattice problem only.
So one can decompose the quasimomentum $k_a$ into two parts
\begin{equation}
\label{tilde} k_a=\tilde k_a +\frac{1}{L}\sum_{b=1}^{M} \lambda_b
\end{equation}
where $\tilde k_a$ does not depend on $\lambda_1,\ldots,\lambda_M$.
We will often use this fact in what follows.

The eigenvalues $E_{N,M}(\{k\})$ of the Hamiltonian (\ref{H}),
\begin{equation} %\label{}
H\ |\Psi_{N,M}(\{k\};\{\lambda\})\rangle= E_{N,M}(\{k\})\
|\Psi_{N,M}(\{k\};\{\lambda\})\rangle ,
\end{equation}
are given as
\begin{equation}
\label{E} E_{N,M}(\{k\})=\sum_{j=1}^{N}(k_j^2-h+B)-2MB.
\end{equation}
In the case considered ($c=\infty$) they depend only on the total
momentum of the auxiliary lattice problem which results in the
additional degeneration of eigenstates in comparison with the case of
finite $c$. In particular, the ground state at zero magnetic field $B$
is degenerate. Indeed, the total
momentum of the auxiliary lattice problem for the ground state should
be equal to zero at $B=0$, and the requirement of the minimal energy
fixes only the total number of particles $N=N_1+N_2$, the numbers
$N_1=N-M$ and $N_2=M$ being otherwise arbitrary (of course, there
exists also the additional degeneracy if even $N$ and $M$ are fixed).
On the contrary, the ground state is not degenerate if $B\ne 0$.
It is filled by the particles of the first kind ($N_1=N$, $N_2=0$) if
$B<0$ and by the particles of the second kind ($N_1=0$, $N_2=N$) if
$B>0$. Thus, there is the phase transition at point $B=0$.
It should be mentioned that the quantization of the total momentum
is the same for the $XXX$ and $XX0$ chains so that the same results
can be obtained using the $XXX$ chain formulation of the auxiliary
lattice problem.

Consider now the properties of the eigenstates of the Hamiltonian.
One can prove that two eigenstates belonging to
different sectors as well as two eigenstates from the same sector but
with different sets of quasimomenta are orthogonal, i.e.,
\begin{equation} %\label{}
\langle\Psi_{N',M'}(\{k'\};\{\lambda'\})|
\Psi_{N,M}(\{k\};\{\lambda\})\rangle=0
\end{equation}
if $N'\ne N$ or $M'\ne M$ or $\{k'\}\ne \{k\}$ or
$\{\lambda'\}\ne \{\lambda\}$
(two sets are considered to be different if
one of them cannot be obtained from the other by a permutation of the
elements). The normalization of eigenstates is easily computed to be
\begin{eqnarray} \label{orth-norm}
\lefteqn{
\langle\Psi_{N,M}(k_1,\ldots,k_N;\lambda_1,\ldots,\lambda_M)
|\Psi_{N,M}(k_1,\ldots,k_N;\lambda_1,\ldots,\lambda_M)\rangle
}
\nonumber\\ &&
=N^M L^N.
\end{eqnarray}
The eigenstates form the complete set,
\begin{equation} \label{complete}
\sum_{N=0}^{\infty}\sum_{M=0}^{N}
\sum_{k_1<\ldots<k_N \atop \lambda_1<\ldots<\lambda_M}^{}
\frac{
|\Psi_{N,M}(\{k\};\{\lambda\})\rangle
\langle\Psi_{N,M}(\{k\};\{\lambda\})|
}{
\langle\Psi_{N,M}(\{k\};\{\lambda\})|
\Psi_{N,M}(\{k\};\{\lambda\})\rangle
}= 1,
\end{equation}
where the summations are performed over all different sets of
solutions  $k_1,\ldots,k_N$ (\ref{k_j}) and
$\lambda_1,\ldots,\lambda_M$ (\ref{lambda_l}) of the Bethe equations
(\ref{Betheeqns}).

Due to completeness of the eigenstates
one can represent, e.g., the
temperature correlation function $G^{(-)}_1(x,t;h,B)$ in the form
\begin{equation} \label{G^T}
G^{(-)}_1(x,t;h,B)
=\frac{
\sum e^{-E_{N,M}(\{k\})/T}
\langle \psi_1^+(x,t) \psi_1(0,0) \rangle_{N,M}
}{\sum e^{-E_{N,M}(\{k\})/T}},
\end{equation}
where $\langle\psi_1^+(x,t) \psi^{}_1(0,0)\rangle_{N,M}$
is the normalized mean value of the operator \linebreak
$\psi_1^+(x,t) \psi^{}_1(0,0)$
with respect to the eigenstate $|\Psi_{N,M}(\{k\};\{\lambda\})\rangle$
and the summations in (\ref{G^T}) are performed as in
(\ref{complete}).  The similar representation can be written for the
correlator $G^{(+)}_1(x,t;h,B)$ as well.

%%%%%%%%%%%%%%%%%%%%%%%%%%%%%%%%%%%%%%%%%%%%%%%%%%%%%%%%%%%%%%%%%%%%%%%
\section{Form factors of local field operators.}

In this Section we calculate form factors of operators
$\psi^+_{\beta}(x,t)$ and $\psi_{\beta}(x,t)$, i.e., their matrix
elements between two eigenstates of the Hamiltonian. Since the form
factors of operators $\psi^+_{\beta}(x,t)$ and $\psi_{\beta}(x,t)$ are
related by means of complex conjugation, it is sufficient to calculate
the form factors  of operators $\psi_{\beta}(x,t)$.  The nonvanishing
form factors of $\psi_{\beta}(x,t)$ are
\begin{equation} \label{cal-F}
{\cal F}^{(\beta)}_{N,M}(x,t)
\equiv \langle\Psi_{N,\tilde M}(\{q\};\{\mu\})
|\psi_{\beta}(x,t)|\Psi_{N+1,M}(\{k\};\{\lambda\})\rangle,
\end{equation}
where the notation
\begin{equation} %\label{}
\tilde M = \left\{
\begin{array}{lll}
M & \mbox{if} & \beta=1 \\
M-1 & \mbox{if} & \beta=2
\end{array}
\right.
\end{equation}
is introduced.

The form factor ${\cal F}^{(\beta)}_{N,M}(x,t)$ depends on the values
of quasimomenta $\{k\},\{\lambda\}$ and $\{q\},\{\mu\}$ corresponding
to the right and the left eigenstates  in (\ref{cal-F}), respectively.
Here and below the quasimomenta $\{k\},\{\lambda\}$ correspond to the
eigenstates in the sector $(N+1,M)$, and the quasimomenta
$\{q\},\{\mu\}$ are prescribed to the eigenstates in the sector
$(N,\tilde M)$.  The Bethe equations for these momenta are
\begin{eqnarray} \label{Bethe-kq}
\lefteqn{
\{k\},\{\lambda\} :
\left\{
\begin{array}{ll}
e^{ik_a L}=\omega (-\e )^N      & \quad a=1,\ldots,N+1 \\
e^{i\lambda_b(N+1)}= (-1)^{M-1} & \quad b=1,\ldots,M
\end{array}
\right.
}
\nonumber\\
\lefteqn{
\{q\},\{\mu\} :
\left\{
\begin{array}{ll}
e^{iq_a L}=\nu (-\e )^{N-1}      & \quad a=1,\ldots,N \\
e^{i\mu_b N} = (-1)^{\tilde M-1} & \quad b=1,\ldots,\tilde M
\end{array}
\right.
}
\end{eqnarray}
where $\omega$ and $\nu$ are eigenvalues of the cyclic shift operators
of the corresponding auxiliary problems:
\begin{equation} \label{omega-nu}
\omega=\exp\left(i\sum_{b=1}^{M}\lambda_b\right), \qquad
\nu=\exp\left(i\sum_{b=1}^{\tilde M}\mu_b\right).
\end{equation}

Using relation (\ref{time}) and the eigenvalues (\ref{E}) of the
Hamiltonian, one extracts the time dependence of the form factor
(\ref{cal-F}):
\begin{equation} \label{time}
{\cal F}^{(\beta)}_{N,M}(x,t)
=\exp\left\{it \biggl(
\sum_{j=1}^{N}q_j^2-\sum_{j=1}^{N+1}k_j^2 +h_{\beta} \biggr)\right\}\
{\cal F}^{(\beta)}_{N,M}(x).
\end{equation}
Here $h_1=h-B$ and $h_2=h+B$ are the chemical potentials of the
particles of type 1 and of type 2, correspondingly. We denote
\begin{equation} %\label{}
{\cal F}^{(\beta)}_{N,M}(x)\equiv
{\cal F}^{(\beta)}_{N,M}(x,0).
\end{equation}
The distance dependence of the form factor is described by function
${\cal F}^{(\beta)}_{N,M}(x)$. Using expressions
(\ref{PsiNM}) for eigenstates one represents it in terms of the wave
functions as
\begin{eqnarray} \label{F(x)}
\lefteqn{
{\cal F}^{(\beta)}_{N,M}(x)
=(N+1)!\int\limits_{0}^{L} dz_1 \ldots \int\limits_{0}^{L} dz_N
\sum_{\alpha_1,\ldots,\alpha_N}^{}
\bar\chi_{N,\tilde M}^{\alpha_1\ldots\alpha_N}
(z_1,\ldots,z_N|\{q\};\{\mu\})
}
\nonumber\\ && \times
\chi_{N+1,M}^{\alpha_1\ldots\alpha_N\beta}
(z_1,\ldots,z_N,x|\{k\};\{\lambda\}).
\end{eqnarray}
Here and below the bar denotes the complex conjugation.
Substituting now the explicit expressions
for the wave functions into (\ref{F(x)}) and producing
the necessary calculation one comes to the expression
\begin{eqnarray} \label{formula}
\lefteqn{
{\cal F}^{(\beta)}_{N,M}(x)
={1 \over N!}
\int\limits_{0}^{L} dz_1 \ldots \int\limits_{0}^{L} dz_N
\sum_{R\in S_N}^{}
\Biggl\{\theta(z_{R_1}<\ldots<z_{R_N}<x) F_\beta(N)
}
\nonumber\\ &&
+\sum_{j=1}^{N-1}
\theta(z_{R_1}<\ldots<z_{R_j}<x<z_{R_{j+1}}<\ldots<z_{R_N})
(-\e )^{N-j} F_\beta(j)
\nonumber\\ &&
+\theta(x<z_{R_1}<\ldots<z_{R_N}) (-\e )^{N} F_\beta(0) \Biggr\}
\sum_{Q\in S_N}^{} (-1)^{[Q]}e^{-i(q_{Q_1}z_1+\cdots+q_{Q_N}z_N)}
\nonumber\\ &&\times
\sum_{P\in S_{N+1}}^{} (-1)^{[P]}
e^{i(k_{P_1}z_1+\cdots+k_{P_N}z_N+k_{P_{N+1}}x)},
\end{eqnarray}
where
\begin{equation}
\label{F_beta} F_\beta(j) \equiv \sum_{\alpha_1,\ldots,\alpha_N}^{}
\bar\xi_{N,\tilde M}^{\alpha_1\ldots\alpha_N}
\xi_{N+1,M}^{\alpha_1\ldots\alpha_j\beta\alpha_{j+1}\ldots\alpha_N}.
\end{equation}

Further derivation requires more detailed explanation.

First, note that $\xi_{N+1,M}$ and $\xi_{N,\tilde M}$ are eigenvectors
of the cyclic shift operators $C_{N+1}$ and $C_{N}$ with eigenvalues
$\omega$ and $\nu$, respectively (see (\ref{C_N}) and (\ref{Bethe-kq}),
(\ref{omega-nu})), namely,
\begin{eqnarray} \label{cicle}
\lefteqn{
\xi^{\alpha_1\alpha_2\ldots\alpha_{N+1}}_{N+1,M}
=\omega\,\xi^{\alpha_2\ldots\alpha_{N+1}\alpha_1}_{N+1,M},
\qquad
\xi^{\alpha_1\alpha_2\ldots\alpha_{N}}_{N,\tilde M}
=\nu\,\xi^{\alpha_2\ldots\alpha_{N}\alpha_1}_{N,\tilde M},
}
\nonumber\\
\lefteqn{
\omega^{N+1}=1,\qquad \nu^{N}=1.
}
\end{eqnarray}
Using relations (\ref{cicle}) $(N-j)$ times for both vectors, one can
move the superscript $\beta$ in (\ref{F_beta}) to the right, preserving
the same order of indexes $\alpha_1,\ldots,\alpha_{N}$ of
$\bar\xi_{N,\tilde M}$ and $\xi_{N+1,M}$ with respect to which the
summation in (\ref{F_beta}) is performed. This results in the
expression
\begin{equation} \label{N-j}
F_\beta(j)
=(\bar\omega \nu)^{N-j} F_\beta, \qquad
F_\beta \equiv F_\beta(N)
\end{equation}
(one should take into account that
$\bar\omega=\omega^{-1},\bar\nu=\nu^{-1}$). Due to (\ref{N-j}),
one can move $F_\beta$ in (\ref{formula}) out of the braces leaving
the factors $(\bar\omega\nu)^{N-j}$ instead of $F_\beta(j)$ in the
first sum (over the permutations $R\in S_N$).

Second, let us introduce the function
\begin{equation} %\label{}
\rho(z)\equiv \theta(z)-\e\bar\omega\nu\theta(-z)
\end{equation}
(which is an analogue of function $\sgn(z)$).
It is not difficult to see that the following relation holds
\begin{eqnarray} \label{sum_R}
\lefteqn{
\sum_{R\in S_N}^{}
\biggl\{\theta(z_{R_1}<\ldots<z_{R_N}<x)
}
\nonumber\\ &&
+\sum_{j=1}^{N-1}
\theta(z_{R_1}<\ldots<z_{R_j}<x<z_{R_{j+1}}<\ldots<z_{R_N})
(-\e\bar\omega\nu)^{N-j}
\nonumber\\ &&
+\theta(x<z_{R_1}<\ldots<z_{R_N}) (-\e\bar\omega\nu)^{N} \biggr\}
\nonumber\\ &&
=\prod_{j=1}^{N}\rho(x-z_j)
\end{eqnarray}
for $z_1, \ldots, z_N, x \in [0,L]$ being all different.  If two of
$z$'s do coincide, then the corresponding terms do not contribute into
(\ref{formula}) due to the fact that the last sum with respect to the
permutations $P\in S_{N+1}$ then vanishes.  Hence the value  $\rho(0)$
is unessential.  Due to (\ref{N-j}) and (\ref{sum_R}), the expression
for ${\cal F}^{(\beta)}_{N,M}(x)$ can be rewritten in the form
\begin{eqnarray} \label{prod-int}
\lefteqn{
{\cal F}^{(\beta)}_{N,M}(x)
=\frac{1}{N!} F_\beta
\int\limits_{0}^{L} dz_1 \ldots \int\limits_{0}^{L} dz_N
\prod_{j=1}^{N}\rho(x-z_j)
}
\nonumber\\ && \times
\sum_{Q\in S_N \atop P\in S_{N+1}}^{} (-1)^{[P]+[Q]}
e^{i(k_{P_1}-q_{Q_1})z_1+\cdots+i(k_{P_N}-q_{Q_N})z_N+ik_{P_{N+1}}x}.
\end{eqnarray}
Since the dependence on $z_1,\ldots,z_N$ in (\ref{prod-int}) is
factorized, the integrals can be taken explicitly,
\begin{equation} \label{integral}
\int\limits_{0}^{L} dz \rho(x-z) e^{i(k-q)z}
= -i (1+\e\bar\omega\nu) \frac{e^{i(k-q)x}}{(k-q)}.
\end{equation}
The Bethe equations (\ref{Bethe-kq}) for $k$ and $q$ were used in
derivation of (\ref{integral}). Due to this relation we have
\begin{eqnarray} \label{f-f}
\lefteqn{
{\cal F}^{(\beta)}_{N,M}(x)
=(-i)^N(1+\e\bar\omega\nu)^N F_\beta
\left\{\sum_{P \in S_{N+1}}^{} (-1)^{[P]}
\frac{1}{k_{P_1}-q_1} \times \cdots \times  \frac{1}{k_{P_N}-q_N}
\right\}
}
\nonumber\\ && \times
\exp\left\{ ix \Biggl( \sum_{j=1}^{N+1}k_j-\sum_{l=1}^{N}q_l \Biggr)
\right\}.
\end{eqnarray}
The expression (\ref{f-f}) is the desired representation for the form
factor ${\cal F}^{(\beta)}_{N,M}(x)$.  As usual, the dependence of the
form factor on $x$ is described by the "translational" exponential.
It is multiplied by the  determinant of a matrix depending on the
momenta $\{k\}$ and $\{q\}$ only
(the sum in the first braces in (\ref{f-f})).
The factor $F_\beta$ depends on the sets $\{\mu\}$ and $\{\lambda\}$ of
the momenta of the auxiliary lattice problems only, being a special
``scalar product'' of two Bethe vectors defined on the lattices with
$N$ and $N+1$ sites:
\begin{equation} \label{F_b}
F_\beta \equiv \sum_{\alpha_1,\ldots,\alpha_N}^{}
\bar\xi_{N,\tilde M}^{\alpha_1\ldots\alpha_N}
\xi_{N+1,M}^{\alpha_1\ldots\alpha_N\beta}.
\end{equation}

Now we are going to obtain the representation for $F_\beta$ as a
determinant of a matrix depending on $\{\mu\}$ and $\{\lambda\}$.
To do this it is suitable to use the notations for the vectors
$\bar\xi_{N,\tilde M}$ and $\xi_{N+1,M}$ as the bra and ket vectors,
$(\xi_{N,\tilde M}|$ and $|\xi_{N+1,M})$, which are built
upon the pseudovacua
$(\Uparrow_N\!|:= \otimes_{n=1}^{N}(\uparrow_n\!|$
and $|\!\Uparrow_{N+1}):= \otimes_{n=1}^{N+1} |\!\uparrow_{n}) $
of spin chains with $N$ and $N+1$ sites, respectively.
Let us introduce the new vector
\begin{equation} %\label{}
(\widetilde\xi_{N,\tilde M}|\equiv (\xi_{N,\tilde M}| \otimes
(\uparrow_{N+1}\!|
\end{equation}
which is built upon the pseudovacuum
$(\Uparrow_{N+1}\!|=\otimes_{n=1}^{N+1} (\uparrow_n\!|$.
Then the scalar product (\ref{F_b}) can be rewritten as
\begin{eqnarray} %\label{}
\lefteqn{
F_1=\Bigl(\widetilde\xi_{N,M}(\{\mu\})\Bigl|\Bigr.
\xi_{N+1,M}(\{\lambda\}\Bigr),
} \label{F_1}
\\
\lefteqn{
F_2=\Bigl(\widetilde\xi_{N,M-1}(\{\mu\})\Bigl|\Bigr.
\sigma^{+}_{(N+1)}\Bigl|\Bigr.
\xi_{N+1,M}(\{\lambda\})\Bigr),
} \label{F_2}
\end{eqnarray}
where the dependence on quasimomenta $\{\mu\}$ and $\{\lambda\}$ is
written explicitly.
Note that there are $M$ quasimomenta in the set $\{\mu\}$ in
(\ref{F_1}) and $M-1$ quasimomenta in the set $\{\mu\}$ in (\ref{F_2});
they satisfy different Bethe equations but nevertheless we denote
these two sets by the same letter. At any particular case, it will be
clear what the Bethe equations for the set $\{\mu\}$ are.

As mentioned above, the factors $F_1$ and $F_2$ given by (\ref{F_1})
and (\ref {F_2}) can be represented as determinants.
This representation will be obtained now for $F_1$. The derivation of
the representation for $F_2$ is quite similar.

Substituting the  expressions for the Bethe vectors involved into
(\ref{F_1}) we get
\begin{eqnarray} \label{F1}
\lefteqn{
F_1 = \sum_{m_1,\ldots,m_M=1}^{N} \sum_{n_1,\ldots,n_M=1}^{N+1}
\overline{\varphi_{N,M}(m_1,\ldots,m_M|\{\mu\})}\,
\varphi_{N+1,M}(n_1,\ldots,n_M|\{\lambda\})
}
\nonumber\\ && \times
\Bigl(
\Uparrow_{N+1}\!\Bigl| \sigma^{+}_{(m_1)} \cdots \sigma^{+}_{(m_M)} \
\sigma^{-}_{(n_1)} \cdots \sigma^{-}_{(n_M)}
\Bigr|\!\Uparrow_{N+1}\Bigr).
\end{eqnarray}
Since there are no $\sigma^{+}_{(N+1)}$ here, all the terms containing
$\sigma^{-}_{(N+1)}$
do not contribute in (\ref{F1}). Hence the summations with respect to
$n_1,\ldots,n_M$ can be performed also from $1$ to $N$. Thus (\ref{F1})
is rewritten as
\begin{eqnarray} %\label{}
\lefteqn{
F_1 = M! \sum_{n_1,\ldots,n_M=1}^{N}
\overline{\varphi_{N,M}(n_1,\ldots,n_M|\{\mu\})}\,
\varphi_{N+1,M}(n_1,\ldots,n_M|\{\lambda\})
}
\nonumber\\ &&
=\frac{1}{M!}
\sum_{n_1,\ldots,n_M=1}^{N}
\sum_{P\in S_M}^{}(-1)^{[P]}e^{-i(\mu_{P_1}n_1+\cdots+\mu_{P_M}n_M)}
\nonumber\\ && \times
\sum_{Q\in S_M}^{}
(-1)^{[Q]} e^{i(\lambda_{Q_1}n_1+\cdots+\lambda_{Q_M}n_M)},
\end{eqnarray}
which gives
\begin{equation} %\label{}
F_1=\sum_{P\in S_M}^{} (-1)^{[P]}
\left(\sum_{n_1=1}^{N}e^{i(\lambda_{P_1}-\mu_1)n_1} \right)
\times\cdots\times \left(\sum_{n_M=1}^{N}e^{i(\lambda_{P_M}-\mu_M)n_M}
\right).
\end{equation}
This is, by definition, the determinant of $M\times M$-dimensional
matrix.  The determinant representation for $F_2$ can be derived in a
similar way, only the terms with $\sigma^{-}_{(N+1)}$ contributing.

Let us now sum up the results of this Section.

For the form factors
${\cal F}^{(\beta)}_{N,M}(x,t)$
we obtain the following representation
\begin{eqnarray} \label{F^1}
\lefteqn{
{\cal F}^{(\beta)}_{N,M}(x,t)
=(-i)^N (1+\e\bar\omega\nu)^N
{\det}_M B_\beta \ {\det}_{N+1} D
}
\nonumber\\ &&\times
\exp\left\{
\sum_{a=1}^{N+1}(-itk_a^2+ixk_a)
-\sum_{a=1}^{N}(-itq_a^2+ixq_a)+ith_\beta
\right\},
\end{eqnarray}
where $h_1=h-B$, $h_2=h+B$. The matrix elements of
$M\times M$-dimensional matrices $B_{1,2}$ are
\begin{eqnarray} \label{B_12}
\lefteqn{
(B_1)_{ab}=\sum_{n=1}^{N} e^{i(\lambda_a-\mu_b)n}, \qquad
a,b=1,\ldots,M,
}
\nonumber\\
\lefteqn{
(B_2)_{ab}=\sum_{n=1}^{N} e^{i(\lambda_a-\mu_b)n},
\quad (B_2)_{a,M}=1,
}
\nonumber\\ &&
a=1,\ldots,M, \quad b=1,\ldots,M-1,
\end{eqnarray}
and $\omega$, $\nu$ are given by (\ref{omega-nu}).
The $(N+1)\times(N+1)$-dimensional matrix $D$ has matrix elements
\begin{equation} \label{D}
(D)_{ab}= \frac{1}{k_a-q_b}, \quad (D)_{a,N+1}=1, \quad
a=1,\ldots,N+1, \quad b=1,\ldots,N.
\end{equation}

It should be noted, that the determinant of the
$(N+1)\times(N+1)$-dimensional matrix $D$ can be represented in terms
of the determinant of the $N\times N$-dimensional matrix:
\begin{equation} \label{D=AA}
{\det}_{N+1}D= \left.\biggl[1+ \frac{\partial}{\partial z}\biggr]
{\det}_N (A_1-z A_2)\right|_{z=0}.
\end{equation}
The $N\times N$-dimensional matrices $A_1$ and $A_2$ have elements
\begin{eqnarray} %\label{}
(A_1)_{ab}=\frac{1}{k_a-q_b},\quad
(A_2)_{ab}=\frac{1}{k_{N+1}-q_b}, \qquad a,b=1,\ldots,N.
\end{eqnarray}
Since the matrix $A_2$ is of rank equal to one, ${\det}_N (A_1-z A_2)$
is a linear function of $z$. Thus, the r.h.s. of (\ref{D=AA}) is equal
to ${\det}_N(A_1-A_2)$. The relation (\ref{D=AA}) will be used in the
next Section.

Let us note that the permitted values of $q_a$ depend on the sum of the
corresponding $\mu$'s.  Hence the values of matrix elements of matrix
$D$ as well as the values of exponents in (\ref{F^1}) are in fact
different in cases $\beta=1$ and $\beta=2$ though being written down in
the same way for the sake of simplicity.

Let us consider the particular cases $M=0$ and $M=N+1$.
If $M=0$ then ${\cal F}^{(2)}_{N,M}(x,t)$ vanishes due to its
definition while ${\cal F}^{(1)}_{N,M}(x,t)$ coincides with the
form factor of the one-component gas. In this case we put
${\det}_{M=0} B_1 \equiv 1$ (see the derivation of $F_1$ above),
$\nu=\omega=1$, and the Bethe equations for $\{k\}$
and $\{q\}$ coincide with those in the one-component gas. Similarly, if
$M=N+1$ then ${\cal F}^{(1)}_{N,M}(x,t)$ vanishes and
${\cal F}^{(2)}_{N,M}(x,t)$
coincides with the form factor of the one-component gas
(in this case $\sum_{b=1}^{N+1}\lambda_b=0$, $\sum_{b=1}^{N}\mu_b=0$,
hence $\nu=\omega=1$ and  Bethe equations for $\{k\}$ and $\{q\}$
coincide with those in the one-component  gas). The value of
${\det}_{M=N+1}B_2$ is a numerical factor related to the
normalization of eigenstates (see (\ref{orth-norm})), being
hence unessential.

%%%%%%%%%%%%%%%%%%%%%%%%%%%%%%%%%%%%%%%%%%%%%%%%%%%%%%%%%%%%%%%%%%%%%%%
\section{Normalized mean values of bilocal operators.}

In this section we derive the representations for
the normalized mean values
\begin{eqnarray} \label{mean-1}
\lefteqn{
\langle\psi_\beta^+(x,t)
\psi^{}_\beta(0,0)\rangle_{N+1,M}
}
\nonumber\\ &&
=\frac{ \langle\Psi_{N+1,M}(\{k\};
\{\lambda\})| \psi^+_\beta(x,t) \psi^{}_\beta(0,0)
|\Psi_{N+1,M}(\{k\};\{\lambda\})\rangle }
{ \langle\Psi_{N+1,M}(\{k\};
\{\lambda\}) |\Psi_{N+1,M}(\{k\};\{\lambda\})\rangle}
\end{eqnarray}
and
\begin{eqnarray} \label{mean-2}
\lefteqn{
\langle\psi^{}_\beta(x,t)
\psi^+_\beta(0,0)\rangle_{N,M}
}
\nonumber\\ &&
=\frac{ \langle\Psi_{N,M}(\{q\};\{\mu\})|
\psi^{}_\beta(x,t) \psi^+_\beta(0,0)
|\Psi_{N,M}(\{q\};\{\mu\})\rangle }{ \langle\Psi_{N,M}(\{q\};\{\mu\})
|\Psi_{N,M}(\{q\};\{\mu\})\rangle}
\end{eqnarray}
($\beta=1,2$) with respect to the eigenstates of the Hamiltonian. These
quantities enter the representations for temperature correlation
functions. Below the derivation of the representation in the case
$\beta=1$ is considered in detail, the derivation in the case $\beta=2$
being quite similar.

Consider the normalized mean value (\ref{mean-1}) for $\beta=1$.
Inserting the complete set of eigenstates (see (\ref{complete}))
between operators $\psi_1^+(x,t)$ and $\psi_1(0,0)$
and taking into account the normalization condition (\ref{orth-norm})
we get
\begin{eqnarray} \label{FF}
\lefteqn{
\langle\psi^+_1(x,t) \psi^{}_1(0,0)\rangle_{N+1,M}
=\frac{1}{L^{2N+1}N^M(N+1)^M}\sum_{\{q\},\{\mu\}}^{}
\overline{{\cal F}^{(1)}_{N,M}(x,t)}\ {\cal F}^{(1)}_{N,M}(0,0)
}
\nonumber\\ &&
=\frac{1}{L^{2N+1} N^M (N+1)^M}
\sum_{\{q\},\{\mu\}}^{}
|1+\e \nu\bar\omega|^{2N}
\left|{\det}_M B_1 \right|^2
({\det}_{N+1} D)^2
\nonumber\\ && \times
\exp\left\{
\sum_{a=1}^{N+1}(itk_a^2-ixk_a)-\sum_{b=1}^{N}(itq_b^2-ixq_b)-ith_1
\right\}.
\end{eqnarray}
Recall that $\omega=\exp i\Lambda$ and $\nu=\exp i\Theta$ where
$\Lambda$ and $\Theta$ are the total momenta of the corresponding
auxiliary lattice problems:
\begin{equation} %\label{}
\Lambda\equiv\sum_{a=1}^{M} \lambda_a,\qquad
\Theta\equiv\sum_{a=1}^{\tilde M} \mu_a.
\end{equation}
The summation over $\{q\},\{\mu\}$
means the summation over all possible sets of
quasimomenta $\{q\}$ and $\{\mu\}$. Since each $q_a$ can be represented
as $q_a =\tilde q_a+\Theta/L$ (see (\ref{tilde}) and the comments
there) where $\tilde q_a$ does not depend on $\mu$'s, one can perform
the summation over $\{\tilde q\}\equiv \tilde q_1,\ldots,\tilde q_N$
independently of the summation over $\{\mu\}$. The expression
under the sum in (\ref{FF}) is symmetric under the permutations of
$q$'s (and $\tilde q$'s as well) and $\mu$'s separately, being
equal to zero whenever two $q$'s or $\mu$'s coincide. Thus one can
change the sum in (\ref{FF}) to the sum over all permitted values of
each $\tilde q_a$ and each $\mu_b$,
\begin{equation} \label{SS}
\sum_{\{q\},\{\mu\}}^{}\equiv
\sum_{q_1<\cdots<q_N\atop \mu_1<\cdots<\mu_M}^{}
\to \frac{1}{N!} \sum_{\tilde q_1}^{}\cdots\sum_{\tilde q_N}^{}\
\frac{1}{M!} \sum_{\mu_1}^{}\cdots\sum_{\mu_M}^{},
\end{equation}
where the sums over each individual
$\tilde q_a\ (a=1,\ldots,N)$ and $\mu_b\ (b=1,\ldots,M)$
are independent:
\begin{equation} \label{sum_q}
\sum_{\tilde q_a}^{} f(q_a)
=\sum_{j\in Z}^{} f((\tilde q_a)_j),\qquad
\sum_{\mu_b}^{} f(\mu_b) =\sum_{l=1}^{N} f((\mu_b)_l).
\end{equation}
The permitted values $(\tilde q_a)_j$ and $(\mu_b)_l$ are solutions of
the Bethe equation (\ref{Bethe-kq}). Explicitly,
\begin{eqnarray} \label{q_j}
\lefteqn{
(\tilde q_a)_j
=\frac{2\pi}{L}\left(-\Bigl(\frac{1+\e}{2}\Bigr)\frac{N-1}{2}+j\right),
\qquad j\in Z,
}
\nonumber\\
\lefteqn{
(\mu_b)_l=
\frac{2\pi}{N}
\left(-\frac{N}{2}-\frac{1+(-1)^{N-M}}{4}+l\right),
\quad l=1,\ldots,N.
}
\end{eqnarray}

Taking into account (\ref{SS}), let us perform first
the summations over $\tilde q_1,\ldots,\tilde q_N$ and then over
$\mu_1,\ldots,\mu_M$ in (\ref{FF}).
Note that the dependence on $q$'s in (\ref{FF}) enters only the
elements of matrix $D$ and the exponential factor.
The determinant of the matrix $D$ can be
represented as the sum over permutations,
\begin{equation} \label{detD}
{\det}_{N+1}D
=\sum_{P\in S_{N=1}}^{} (-1)^{[P]} \prod_{a=1}^{N}
\frac{1}{k_{P_a}-q_a},
\end{equation}
so that
the dependence on each $q_a$ in the r.h.s. of (\ref{FF}) is factorized.
Therefore we can use the technique similar to that applied in the cases
of the one-component impenetrable Bose-gas \cite{KS-90,KS-91} and
of $XX0$ model \cite{CIKT-92,CIKT-93}. This procedure (which can be
called ``inserting the summation into determinant'') results in our
case in the following indentity:
\begin{eqnarray} \label{ins-q}
\lefteqn{
\frac{|1+\e \nu\bar\omega|^{2N}}{L^{2N+1}N!}
\sum_{\tilde q_1}^{}\cdots \sum_{\tilde q_N}^{} ({\det}_{N+1} D)^2
\exp\left\{
\sum_{a=1}^{N+1}(itk_a^2-ixk_a)-\sum_{b=1}^{N}(itq_b^2-ixq_b)
\right\}
}
\nonumber\\ &&
=\left.
\frac{\partial}{\partial z}{\det}_{N+1}(S^{(-)}+zR^{(-)})
\right|_{z=0}.
\end{eqnarray}
The matrix elements of the
$(N+1)\times(N+1)$ matrices $S^{(-)}$ and $R^{(-)}$ are
\begin{eqnarray} \label{S}
\lefteqn{
(S^{(-)})_{ab}
=e_{-}(k_a)\ e_{-}(k_b)\
\frac{|1+\e \nu\bar\omega|^{2}}{L^2}\
\sum_{\tilde q}^{} \frac{e^{-itq^2+ixq}}{(k_a-q)(k_b-q)},
\qquad q=\tilde q + \frac{\Theta}{L},
}
\nonumber\\
\lefteqn{
(R^{(-)})_{ab} = \frac{e_{-}(k_a)e_{-}(k_b)}{L},
}
\end{eqnarray}
the function $e_{-}(k_a)$ is defined as
\begin{equation} %\label{}
e_{-}(k_a)=\exp\left(\frac{itk_a^2-ixk_a}{2}\right)
\end{equation}
and $z$ is a parameter. Since the rank of matrix $R^{(-)}$ is equal to
one, ${\det}_{N+1} (S^{(-)}+zR^{(-)})$ is a linear function of $z$.

To go to the thermodynamic limit (which is done in the next Section)
it is necessary to rewrite the expressions for the matrix elements of
the matrix $S^{(-)}$. Note that matrix elements given by (\ref{S})
for finite $L$ are well-defined functions, since all possible zeros
of the denominator of the expression under the sum over $\tilde q$
are canceled by zeros of the numerator coming from the factor
$|1+\e \nu\bar\omega|^{2}$.
It is due to the fact that the
condition $k=q$ can be satisfied only
if $\Lambda-\Theta=\pi\frac{1+\e}{2} (\mod 2\pi)$, which follows from
the Bethe equations.
In the limit $L\to \infty$, however, the values of $k$ and $q$
become arbitrary, the summation over $\tilde q$
is changed for
the  integration $\frac{1}{L}\sum_{\tilde q}^{} \to
\frac{1}{2\pi} \int_{-\infty}^{\infty} d\tilde q$ and poles on
the integration contour appear which should be taken into account
before going to the limit $L\to\infty$. Therefore one
has to rewrite the matrix elements of $S^{(-)}$ in terms of
functions which are well defined in the thermodynamic limit.

The necessary calculations are given in Appendix. As the result,
the matrix $S^{(-)}$ can be represented in the form
\begin{equation} \label{IVQ}
S^{(-)}
=I+\frac{1+\e\,\cos(\Lambda-\Theta)}{2}\ V^{(-)}_1
+\frac{\e\,\sin(\Lambda-\Theta)}{2}\ V^{(-)}_2,
\end{equation}
where $I$ is the unit $(N+1)\times (N+1)$ matrix,
$(I)_{ab}=\delta_{ab}$, and matrices $V^{(-)}_{1,2}$ are
\begin{eqnarray} \label{VQ}
\lefteqn{
( V^{(-)}_1)_{ab}
=\frac{2}{L}\,
\frac{e^{(-)}_{+}(k_a) e^{}_{-}(k_b)-e^{}_{-}(k_a) e^{(-)}_{+}(k_b)}
{k_a-k_b},
}
\nonumber\\
\lefteqn{
( V^{(-)}_2)_{ab}
=\frac{2}{L}
\frac{[e_{-}(k_a)]^{-1} e_{-}(k_b)-e_{-}(k_a) [e_{-}(k_b)]^{-1}}
{k_a-k_b}.
}
\end{eqnarray}
The functions involved are defined as
\begin{equation} %\label{}
e^{(-)}_{+}(k)=e^{(-)}(k) e^{}_{-}(k), \qquad
e^{}_{-}(k)=e^{\frac{itk^2-ixk}{2}},
\end{equation}
\begin{equation} \label{ed}
e^{(-)}(k)
=\frac{2}{L}\sum_{\tilde q}^{}
\frac{e^{-itq^2+ixq}-e^{-itk^2+ixk}}{q-k},
\qquad q=\tilde q+\frac{\Theta}{L}.
\end{equation}
These functions are well defined in the thermodynamic limit.

The summation over $\tilde q$'s being fulfilled (\ref{ins-q}),
one remains
with the sum over the quasimomenta $\mu$'s in the expression
(\ref{FF}) for the normalized mean value:
\begin{eqnarray} \label{3det}
\lefteqn{
\langle\psi^+_1(x,t) \psi^{}_1(0,0)\rangle_{N+1,M}
=\frac{1}{N^M (N+1)^M M!} \sum_{\mu_1}^{}\cdots \sum_{\mu_M}^{}
\left| {\det}_M B_1\right|^2
}
\nonumber\\ && \times
e^{-ith_1}
\left.\frac{\partial}{\partial z} {\det}_{N+1}(S^{(-)}+zR^{(-)})
\right|_{z=0}
\end{eqnarray}

Let us perform the summation over $\mu_1,\ldots,\mu_M$.
Note that each element of matrix $S^{(-)}$ depends on
the sum $ \Theta\equiv\mu_1+\cdots+\mu_M$ only. Moreover,
since the summation over $\tilde q$ in (\ref{S}) is governed by the
rules (\ref{sum_q}), (\ref{q_j}) and since $q=\tilde q+\Theta/L$,
matrix elements
$(S)_{ab}$ are periodic functions of $\Theta$ with the period $2\pi$.
Matrix $R^{(-)}$ does not depend on $ \mu_1,\ldots,\mu_M$ at
all, hence ${\det}_{N+1}(S^{(-)}+zR^{(-)})$ is also a periodic function
of $\Theta$. Further, $\exp i\Theta = \nu$ (see (\ref{omega-nu}))
is an eigenvalue of cyclic shift operator $C_N$, hence
$\Theta$ takes the values $2\pi n/N$ where $n=0,1,\ldots,N-1\ (\mod N)$
and we can write
\begin{equation} \label{unit}
1=\sum_{n=0}^{N-1}
\delta_{(N)}\left(N\frac{\mu_1+\cdots+\mu_M}{2\pi}-n\right),
\end{equation}
where $\delta_{(N)}$ is Kronecker symbol on $Z_N$, defined as
\begin{equation} %\label{}
m\in Z,\quad \delta_{(N)}(m) = \left\{
\begin{array}{ll}
1 & \mbox{if}\ m=0 \ (\mod N) \\ 0 & \mbox{otherwise}
\end{array}
\right.
\end{equation}
and can be represented as Fourier sum
\begin{equation} \label{delta}
\delta_{(N)}(m)
=\frac{1}{N}\sum_{p=0}^{N-1} e^{\frac{2\pi i}{N}pm}.
\end{equation}
Inserting (\ref{unit})  under the sum over
$\mu_1,\ldots,\mu_M$ in (\ref{3det})
we can change $\Theta\to 2\pi n/N$ everywhere
in the expressions for elements of the matrix $S^{(-)}$.
Using the representation (\ref{delta}) we can rewrite (\ref{3det}) as
\begin{eqnarray} \label{e(mu)}
\lefteqn{
\langle\psi^+_1(x,t) \psi^{}_1(0,0)\rangle_{N+1,M}
}
\nonumber\\ &&
=\frac{1}{N^{M+1} (N+1)^M M!}
\sum_{\mu_1}^{}\cdots \sum_{\mu_M}^{}
\sum_{n,p=0}^{N-1} e^{ip(\mu_1+\cdots+\mu_M)-\frac{2\pi i}{N}pn}\
\left| {\det}_M B_1 \right|^2
\nonumber\\ && \times
e^{-ith_1}
\left.\frac{\partial}{\partial z} {\det}_{N+1}(S^{(-)}_n+zR^{(-)})
\right|_{z=0},
\end{eqnarray}
where matrices $S^{(-)}_n$ now do not depend on $\mu_1,\ldots,\mu_M$ at
all and their elements are
$(S^{(-)}_n)_{ab}=(S^{(-)})_{ab}|_{\Theta=2\pi n/N}$. Matrices
$S^{(-)}_n$ have the structure corresponding to (\ref{IVQ}),
\begin{equation} \label{IVnQ}
S^{(-)}_{n}
=I+\frac{1+\e\,\cos(\Lambda-\frac{2\pi n}{N})}{2} V^{(-)}_{1,n}
+\frac{\e\,\sin(\Lambda-\frac{2\pi n}{N})}{2} V^{(-)}_2,
\end{equation}
where $V^{(-)}_{1,n}= V^{(-)}_1|_{\Theta=2\pi n/N}$, i.e., one should
put $q=\tilde q + \frac{2\pi n}{NL}$
in the function $e^{(-)}$ (\ref{ed})
determining the elements of the matrix
$V^{(-)}_1$.

Considering the determinant of matrix $B_1$
as the sum over permutations,
\begin{equation} %\label{}
{\det}_M B_1 = \sum_{P\in S_M}^{} (-1)^{[P]} \prod_{a=1}^{M}
\left[\sum_{n_a=1}^{N}e^{i(\lambda_{P_a}-\mu_a)n_a} \right],
\end{equation}
we see that the dependence on
$\mu_1,\ldots,\mu_M$ in (\ref{e(mu)}) can be factorized.
Thus we can
perform the procedure of ``inserting summation into determinant'' with
respect to the summations over $\mu_1,\ldots,\mu_M$. Picking up the
relevant terms from (\ref{e(mu)}) one comes to the indentity
\begin{equation} \label{ins-mu1}
\frac{1}{N^M (N+1)^M M!} \sum_{\mu_1}^{}\cdots \sum_{\mu_M}^{}
e^{ip(\mu_1+\cdots+\mu_M)}
\left| {\det}_M  B_1  \right|^2
= {\det}_M U^{(1,-)}_p,
\end{equation}
where $M\times M$ matrices $U^{(1,-)}_p$ have elements
\begin{equation} \label{Up-}
(U^{(1,-)}_p)_{ab} = \frac{1}{N(N+1)}
\sum_{n=1}^{N}\sum_{m=1}^{N}\sum_{\mu}^{}
e^{i(p+m-n)\mu+in\lambda_{a}-im\lambda_{b}}.
\end{equation}

Finally, the normalized mean value of operator
$\psi^+_1(x,t)\psi^{}_1(0,0)$ takes the following form
\begin{eqnarray} \label{NMV_1}
\lefteqn{
\langle\psi^+_1(x,t) \psi^{}_1(0,0)\rangle_{N+1,M}
=e^{-ith_1}\frac{1}{N}\sum_{n,p=0}^{N-1}
e^{-\frac{2\pi i}{N}pn}
}
\nonumber\\ && \times
{\det}_M U^{(1,-)}_p
\left[{\det}_{N+1}(S^{(-)}_n+R^{(-)})-{\det}_{N+1}S^{(-)}_n\right],
\end{eqnarray}
where the property $\frac{\partial}{\partial z} f(z)|_{z=0}=f(1)-f(0)$
valid for a linear function $f(z)$ is used.
Expression (\ref{NMV_1}) is the final expression for the normalized
mean value considered.

For the normalized mean value
$\langle\psi^+_2(x,t)\psi^{}_2(0,0)\rangle_{N+1,M}$ all the
calculations presented above can be done similarly.
The main difference is in the summation over the
values of quasimomenta of the
auxiliary lattice problem, i.e., over the set $\{\mu\}$, which in the
latter case is $\{\mu\}=\mu_1,\ldots,\mu_{M-1}$.
The representation analogous to equation (\ref{NMV_1}) for this
normalized mean value is
\begin{eqnarray} \label{NMV_2}
\lefteqn{
\langle\psi^+_2(x,t) \psi^{}_2(0,0)\rangle_{N+1,M}
=e^{-ith_2}\frac{1}{N}\sum_{n,p=0}^{N-1}
e^{-\frac{2\pi i}{N}pn}
}
\nonumber\\ && \times
\left[{\det}_M (U^{(2,-)}_p+P^{(-)})-{\det}_M U^{(2,-)}_p \right]
\nonumber\\ && \times
\left[{\det}_{N+1}(S^{(-)}_n+R^{(-)})-{\det}_{N+1}S^{(-)}_n\right].
\end{eqnarray}
Here the $M\times M$ matrix $P^{(-)}$ is of rank one
\begin{equation} %\label{}
(P^{(-)})_{ab}=\frac{1}{N+1}
\end{equation}
and matrices $U^{(2,-)}_p$ have the elements
\begin{equation} %\label{}
(U^{(2,-)}_p)_{ab}=\frac{1}{N(N+1)} \sum_{n=1}^{N}\sum_{m=1}^{N}
\sum_{\mu}^{} e^{i(p+m-n)\mu+in\lambda_{a}-im\lambda_{b}}.
\end{equation}
It should be noted that
though the matrices $U^{(2,-)}_p$ and $U^{(1,-)}_p$ look formally the
same, their matrix elements are different (except, e.g., the case
$p=0$, see below) since quasimomenta
$\{\mu\}$ in these two cases satisfy different
Bethe equations.

Consider the particular case $x=0$, $t=0$. Then $S_n=I$
and only the terms with $p=0$ contribute into the
normalized mean values. It is easy to see that
\begin{equation} %\label{}
(U_0^{(1,-)})_{ab}=(U_0^{(2,-)})_{ab}=\delta_{ab}-\frac{1}{N+1}
\end{equation}
and it follows from (\ref{NMV_1}), (\ref{NMV_2}) that
\begin{eqnarray} %\label{}
\lefteqn{
\langle\psi^+_1(0,0) \psi^{}_1(0,0)\rangle_{N+1,M}
=\frac{(N+1)-M}{L},
}
\nonumber\\
\lefteqn{
\langle\psi^+_2(0,0) \psi^{}_2(0,0)\rangle_{N+1,M}
=\frac{M}{L},
}
\end{eqnarray}
which is quite obvious result since in this case the
normalized mean values are just the expectation values of density
operators for particles of type~1 and type~2, respectively.

Let us consider now another normalized mean value, given by
(\ref{mean-2}). Inserting the complete set of eigenstates we get
\begin{eqnarray} \label{FF2}
\lefteqn{
\langle\psi^{}_1(x,t) \psi^+_1(0,0)\rangle_{N,M}
=\frac{1}{L^{2N+1} N^M (N+1)^M}\sum_{\{k\},\{\lambda\}}^{}
{\cal F}^{(1)}_{N,M}(x,t)\ \overline{{\cal F}^{(1)}_{N,M}(0,0)}
}
\nonumber\\ &&
=\frac{1}{L^{2N+1} N^M (N+1)^M}
\sum_{\{k\},\{\lambda\}}^{}
|1+\e \nu\bar\omega|^{2N}
\left|{\det}_M B_1\right|^2
\nonumber\\ && \times
{\det}_{N+1} D\
\left.
\biggl[1+\frac{\partial}{\partial z}\biggr]{\det}_N (A_1-zA_2)
\right|_{z=0}
\nonumber\\ && \times
\exp\left\{\sum_{b=1}^{N}(itq_b^2-ixq_b)
-\sum_{a=1}^{N+1}(itk_a^2-ixk_a)+ith_1
\right\},
\end{eqnarray}
where the relation (\ref{D=AA}) is used for one of the form factors.
Again, as above, we can perform the summation as follows
\begin{equation} \label{SS2}
\sum_{\{k\},\{\lambda\}}^{}\equiv
\sum_{k_1<\cdots<k_{N+1}\atop\lambda_1<\cdots<\lambda_M}^{}
\to \frac{1}{(N+1)!}
\sum_{\tilde k_1}^{} \cdots\sum_{\tilde k_{N+1}}^{}\ \frac{1}{M!}
\sum_{\lambda_1}^{}\cdots \sum_{\lambda_M}^{}.
\end{equation}
The procedure of ``inserting the summation under determinant'' with
respect to the summation over $\tilde k_1,\ldots,\tilde k_{N+1}$ gives
the following result:
\begin{eqnarray} \label{ins-k}
\lefteqn{
\frac{
|1+\e \nu\bar\omega|^{2N}
}{L^{2N+1}(N+1)!}
\sum_{\tilde k_1}^{}\cdots \sum_{\tilde k_{N+1}}^{}
{\det}_{N+1} D
\left.
\biggl[1+\frac{\partial}{\partial z}\biggr]{\det}_N (A_1-zA_2)
\right|_{z=0}
}
\nonumber\\ && \times
\exp\left\{\sum_{b=1}^{N}(itq_b^2-ixq_b)
-\sum_{a=1}^{N+1}(itk_a^2-ixk_a)\right\}
\nonumber\\ &&
= \left.
\biggl[g(x,t)+\frac{\partial}{\partial z}\biggr]{\det}_N
(S^{(+)}-zR^{(+)})
\right|_{z=0}.
\end{eqnarray}
Here we introduce the function
\begin{equation} \label{g(x,t)}
g(x,t) = \frac{1}{L}\sum_{\tilde k}^{} e^{-itk^2+ixk},
\qquad k=\tilde k+\frac{\Lambda}{L}.
\end{equation}
The $N\times N$ matrices $S^{(+)}$ and $R^{(+)}$ are
\begin{eqnarray} \label{SR}
\lefteqn{
(S^{(+)})_{ab}
=e_{-}(q_a)\ e_{-}(q_b)\
\frac{|1+\e \nu\bar\omega|^{2}}{L^2}
\sum_{\tilde k}^{} \frac{e^{-itk^2+ixk}}{(k-q_a)(k-q_b)},
\qquad k =\tilde k + \frac{\Lambda}{L},
}
\nonumber\\
\lefteqn{
(R^{(+)})_{ab}
=\frac{|1+\e \nu\bar\omega|^{2}}{L^3}
\left[e_{-}(q_a) \sum_{\tilde k}^{}
\frac{e^{-itk^2+ixk}}{k-q_a}\right]
\left[e_{-}(q_b) \sum_{\tilde k'}^{}
\frac{e^{-itk'^2+ixk'}}{k'-q_b}\right].
}
\end{eqnarray}
The matrix $R^{(+)}$ is of rank equal to one, and ${\det}_N
(S^{(+)}-zR^{(+)})$ is a linear function of the parameter $z$.

These matrices can be put into the form (see Appendix)
\begin{eqnarray} \label{S+R+}
\lefteqn{
S^{(+)}=I+
\frac{1+\e\,\cos(\Lambda-\Theta)}{2}\ V^{(+)}_1
-\frac{\e\,\sin(\Lambda-\Theta)}{2}\ V^{(+)}_2,
}
\nonumber\\
\lefteqn{
R^{(+)}
=\frac{1+\e\,\cos(\Lambda-\Theta)}{2}\  R^{(+)}_1
}
\nonumber\\ &&
-\e\,\sin(\Lambda-\Theta)\   R^{(+)}_2
+\frac{1-\e\,\cos(\Lambda-\Theta)}{2}\ R^{(+)}_3,
\end{eqnarray}
where matrices involved are
\begin{eqnarray} \label{VR}
\lefteqn{
(V^{(+)}_1)_{ab}
=\frac{2}{L}\,
\frac{e^{(+)}_{+}(q_a)e^{}_{-}(q_b)-e^{}_{-}(q_a)e^{(+)}_{+}(q_b)}
{q_a-q_b},
}
\nonumber\\
\lefteqn{
(V^{(+)}_2)_{ab}
=\frac{2}{L}\,
\frac{[e_{-}(q_a)]^{-1} e_{-}(q_b)-e_{-}(q_a) [e_{-}(q_b)]^{-1}
}{q_a-q_b},
}
\nonumber\\
\lefteqn{
(R^{(+)}_1)_{ab}
=\frac{e^{(+)}_{+}(q_a) e^{(+)}_{+}(q_b)}{L},
}
\nonumber\\
\lefteqn{
(R^{(+)}_2)_{ab}
=\frac{1}{2L}\left[
\frac{e^{(+)}_{+}(q_a)}{e_{-}(q_b)}+\frac{e^{(+)}_{+}(q_b)}{e_{-}(q_a)}
\right],
}
\nonumber\\
\lefteqn{
(R^{(+)}_3)_{ab}
=\frac{1}{L e_{-}(q_a) e_{-}(q_b)}.
}
\end{eqnarray}
Here
\begin{equation} %\label{}
e^{(+)}_{+}(q)= e_{-}(q) e^{+}(q), \quad
e_{-}(q)=e^{\frac{itq^2-ixq}{2}}
\end{equation}
and
\begin{equation} %\label{}
e^{(+)}(q)
=\frac{2}{L} \sum_{\tilde k}^{}
\frac{e^{-itk^2+ixk}-e^{-itq^2+ixq}}{k-q}, \quad
k=\tilde k+\frac{\Lambda}{L}.
\end{equation}

Due to (\ref{ins-k}), the normalized mean value (\ref{FF2})
can be rewritten as
\begin{eqnarray} \label{3det2}
\lefteqn{
\langle\psi^{}_1(x,t) \psi^+_1(0,0)\rangle_{N,M}
=\frac{1}{N^M (N+1)^M M!}
\sum_{\lambda_1}^{}\cdots \sum_{\lambda_M}^{}
\left|{\det}_M B_1 \right|^2
}
\nonumber\\ && \times
e^{ith_1}
\left.
\biggl[g(x,t)+\frac{\partial}{\partial z}\biggr]
{\det}_N(S^{(+)}-zR^{(+)})
\right|_{z=0}.
\end{eqnarray}
The procedure of ``inserting the summation into determinant''
with respect to $\lambda$'s is performed in a
similar way as above.
Using the fact that ${\det}_{N+1}(S^{(+)}-zR^{(+)})$
is a $2\pi$ periodic function of $\Lambda$ and
that $\Lambda=\frac{2\pi m}{N+1}$ where $m=0,1,\ldots,N\ (\mod N+1)$ we
insert
\begin{equation} \label{KronLambda}
1=\sum_{m=0}^{N}
\delta_{(N+1)}\Bigl(m-
(N+1)\frac{\lambda_1+\cdots+\lambda_M}{2\pi}\Bigr)
\end{equation}
under the sums over $\lambda$'s in (\ref{3det2}). Changing
$\Lambda\to\frac{2\pi m}{N+1}$ in the elements of matrices
$S^{(+)}, R^{(+)}$ and in function $g(x,t)$ (\ref{g(x,t)})
and using the representation
(\ref {delta}) for the Kronecker symbol, we rewrite (\ref{3det2}) as
\begin{eqnarray} \label{e(lambda)}
\lefteqn{
\langle\psi^{}_1(x,t) \psi^+_1(0,0)\rangle_{N,M}
}
\nonumber\\ &&
=\frac{1}{N^M (N+1)^{M+1} M!}
\sum_{\lambda_1}^{}\cdots \sum_{\lambda_M}^{}
\sum_{r,m=0}^{N}
e^{\frac{2\pi i}{N+1}rm-ir(\lambda_1+\cdots+\lambda_M)}\
\left|{\det}_M B_1 \right|^2
\nonumber\\ && \times
e^{ith_1}
\left.\biggl[g_m(x,t)+\frac{\partial}{\partial z}\biggr]
{\det}_N (S_m^{(+)}-zR_m^{(+)})
\right|_{z=0},
\end{eqnarray}
where subscript $m$ means that everywhere in the functions determining
the elements of the matrices $S^{(+)},R^{(+)}$ and in the function
$g(x,t)$ one should put $2\pi m/(N+1)$ instead of $\Lambda$.

Similarly to (\ref{ins-mu1}), we have the relation
\begin{equation} %\label{}
\frac{1}{N^M (N+1)^M M!}
\sum_{\lambda_1}^{}\cdots \sum_{\lambda_M}^{}
e^{-ir(\lambda_1+\cdots+\lambda_M)}
\left|{\det}_M B_1\right|^2
= {\det}_M U^{(1,+)}_r,
\end{equation}
where the $M\times M$ matrices $U^{(1,+)}_r$ have elements
\begin{equation} \label{Up+}
(U^{(1,+)}_r)_{ab}
= \frac{1}{N(N+1)} \sum_{n=1}^{N}\sum_{m=1}^{N}\sum_{\lambda}^{}
e^{-i(r+m-n)\lambda+in\mu_{a}-im\mu_{b}}.
\end{equation}

Finally, for the normalized mean value
$\langle\psi^{}_1(x,t)\psi^+_1(0,0)\rangle_{N,M}$ we obtain the
representation
\begin{eqnarray} \label{NMV+}
\lefteqn{
\langle\psi^{}_1(x,t) \psi^+_1(0,0)\rangle_{N,M}
=e^{ith_1} \frac{1}{N+1}\sum_{r,m=0}^{N}
e^{\frac{2\pi i}{N+1}rm}
}
\nonumber\\ && \times
{\det}_M U^{(1,+)}_r
\left[
{\det}_N (S_m^{(+)}-R_m^{(+)})+(g_m(x,t)-1)\,{\det}_N S_m^{(+)}
\right].
\end{eqnarray}

For the sake of completeness, let us write down the representation for
normalized mean value of operator $\psi^{}_2(x,t)\psi^+_2(0,0)$:
\begin{eqnarray} \label{NMV+2}
\lefteqn{
\langle\psi^{}_2(x,t) \psi^+_2(0,0)\rangle_{N,M-1}
=e^{ith_1} \frac{1}{N+1} \sum_{r,m=0}^{N}
e^{\frac{2\pi i}{N+1}rm}\
}
\nonumber\\ && \times
\left[{\det}_{M-1} (U^{(2,+)}_r-P^{(+)}_r)
+(\delta_{r,0}-1)\,{\det}_{M-1} U^{(2,+)}_r\right]
\nonumber\\ && \times
\left[
{\det}_N(S_m^{(+)}-R_m^{(+)})+(g_m(x,t)-1)\,{\det}_N S_m^{(+)}
\right].
\end{eqnarray}
Here the $(M-1)\times(M-1)$ matrices $U^{(2,+)}_r$ have elements
(recall that in this case there are $M-1$ $\mu$'s in the set $\{\mu\}$)
\begin{equation} %\label{}
(U^{(2,+)}_r)_{ab}
= \frac{1}{N(N+1)} \sum_{n=1}^{N}\sum_{m=1}^{N}\sum_{\lambda}^{}
e^{-i(r+m-n)\lambda+im\mu_{a}-in\mu_{b}}
\end{equation}
and the matrix $P^{(+)}_r$ is of rank equal to one
\begin{equation} %\label{}
(P^{(+)}_r)_{ab} = \frac{1}{N(N+1)^2}
\sum_{\lambda}^{} \sum_{m=1}^{N}
e^{-ir\lambda-im(\lambda-\mu_a)}
\sum_{\lambda^{\prime}}^{} \sum_{n=1}^{N}
e^{-ir\lambda^{\prime}+in(\lambda^{\prime}-\mu_b)}.
\end{equation}

The results obtained in this Section allow us to calculate the
correlation functions in the thermodynamic limit.

%%%%%%%%%%%%%%%%%%%%%%%%%%%%%%%%%%%%%%%%%%%%%%%%%%%%%%%%%%%%%%%%%%%%%%%
\section{Temperature correlators in the thermodynamic limit}

Here we present the derivation of
the temperature correlation functions
$G_{1,2}^{(\pm)}$ (\ref{T-cor})
in the thermodynamic limit $L\to\infty$. The
correlation functions $G_1^{(\pm)}$ and $G_2^{(\pm)}$ are related by
inversing the sign of the external field,
\begin{equation} \label{2-1}
G_2^{(\pm)}(x,t;h,B)= G_1^{(\pm)}(x,t;h,-B),
\end{equation}
so that it is sufficient to calculate only one of them
(an independent calculation of both correlators confirm, of course,
the relation (\ref{2-1})). Our result is the determinant representation
of the correlators which is the generalization of the results
obtained earlier for one-component models
\cite{KS-90,KS-91,CIKT-92,CIKT-93}.

Consider first the partition function of the gas which is
the denominator in (\ref{T-cor}). It can be written in the form
\begin{equation} \label{form}
\Tr\left[e^{-H/T}\right]
=1+\sum_{N=1}^{\infty} \sum_{M=0}^{N}
\sum_{\tilde q_1<\ldots<\tilde q_N}^{}
\sum_{\mu_1<\ldots<\mu_M}^{} e^{-E_{N,M}(\{q\})/T},
\end{equation}
where the energy in the sector $(N,M)$ is given by
\begin{equation} %\label{}
E_{N,M}(\{q\})=\sum_{a=1}^{N}(q_a^2-h+B)-2MB.
\end{equation}
As usual,
\begin{equation} \label{qqt}
q_a=\tilde q_a +\frac{\Theta}{L},\qquad
\Theta=\mu_1+\cdots+\mu_M.
\end{equation}
For finite $L$, the partition function can be presented as an
explicit linear combination of several determinants of ``discrete''
infinite dimensional matrices. This representation allows us to
obtain strictly the thermodynamic limit ($L\to \infty$) of the
partition function and also calculate the $1/L$ corrections to the
free energy. The corresponding results will be given in a separate
publication. Below we will consider the thermodynamic limit only.

Note that due to the summation over all possible
values of $\tilde q_1<\ldots<\tilde q_N$ in (\ref{form}) it is
possible, by making the same shift of all $\tilde q_a$,
$\tilde q_a\to \tilde q'_a$, to present each $q_a$ as
\begin{equation} %\label{}
q_a= \tilde q'_a + {\tilde\Theta \over L},\qquad
|\tilde\Theta|\leq{\pi \over L}.
\end{equation}
So in the thermodynamic limit one can neglect $\Theta$ in (\ref{qqt}).
For the same reason one can neglect in the limit also the difference
between the permitted values of $\tilde q_a$ in the sectors with odd or
even number of particles $N$ (see (\ref{k_j})). Under this conditions
the summation over $\mu_1<\ldots<\mu_M$ and over $M$ can be done
with the result
\begin{equation} %\label{}
\sum_{M=0}^{N}\sum_{\mu_1<\ldots<\mu_M}^{} e^{\frac{2B}{T}M}
= (1+e^{\frac{2B}{T}})^N.
\end{equation}
The permitted values of quasimomenta $\tilde q_a\ (a=1,\ldots,N)$,
\begin{equation} %\label{}
(\tilde q_a)_{j+1}-(\tilde q_a)_j =\frac{2\pi}{L},
\end{equation}
fill the interval $(-\infty,\infty)$ densely in the thermodynamic limit
and the sum with respect to $\tilde q_a$ should be changed for the
integral
\begin{equation} \label{s-i}
\frac{1}{L}\sum_{\tilde q}^{}f(\tilde q)
=\frac{1}{L}\sum_{j\in Z}^{} f((\tilde q)_j)
\longrightarrow \frac{1}{2\pi}\int\limits_{-\infty}^{\infty}d\tilde q
f(\tilde q).
\end{equation}
Thus, we have for the partition function
\begin{eqnarray} \label{partition}
\lefteqn{
\Tr\left[e^{-H/T}\right]=1+\sum_{N=1}^{\infty}
\sum_{\tilde q_1<\ldots<\tilde q_N}^{}
(1+e^{\frac{2B}{T}})^N
e^{-\frac{1}{T}\sum_{a=1}^{N}(q_a^2-h+B)}
}
\nonumber\\ &&
=\prod_{\tilde q} \left[
1+ 2 \cosh\!\matrix{\frac{B}{T}} e^{-\frac{q^2-h}{T}}
\right]
\nonumber\\ &&
\cong
\exp\left\{\frac{L}{2\pi}
\int\limits_{-\infty}^{\infty} dq\,
\ln\Biggl(1+ 2 \cosh\!\matrix{\frac{B}{T}} e^{-\frac{q^2-h}{T}}
\Biggr)
\right\},\qquad
L\to\infty.
\end{eqnarray}

Our purpose in what follows is to present the correlators as Fredholm
determinants of linear integral operators. The typical expression for
both the numerator and the denominator (the partition function)
in (\ref{T-cor}) is of the form
\begin{equation} \label{FA}
F=\sum_{N=0}^{\infty} \sum_{q_1<\ldots<q_N}^{}
{\det}_{N} A
=\sum_{N=0}^{\infty} \sum_{q_1<\ldots<q_N}^{}
\left|
\matrix{A(q_1,q_1)& \cdots & A(q_1,q_N)\cr
\vdots & \ddots & \vdots \cr
A(q_N,q_1)& \cdots & A(q_N,q_N)}
\right|
\end{equation}
(by definition, ${\det}_0 A=1$ for any matrix $A$).
The sum in (\ref{FA}) is taken over all permitted values of
$q_1,\ldots,q_N$. In the thermodynamic limit, as explained above,
one can neglect the difference between the permitted values of $q$'s
in the sector $(N,M)$ and put, e.g., $(q)_j=\frac{2\pi}{L}j$
(in other words, it is essential only that
$(q)_{j+1}-(q)_j=\frac{2\pi}{L}$). Then,  due to (\ref{s-i}),
in the limit $L\to\infty$ the expansion
(\ref{FA}) is just the expansion for the Fredholm
determinant,
\begin{equation} \label{Fredholm}
F=\sum_{N=0}^{\infty} \int\limits_{q_1<\ldots<q_N}^{}
d q_1\cdots d q_N\,
{\det}_N {\cal A}
=\det(\hat I+\hat{\cal A}).
\end{equation}
Here, by definition, we denote ${\det}_N {\cal A}$ the determinants
of the $N\times N$--dimensional matrices with matrix elements
${\cal A}(q_a,q_b)$:
\begin{equation} %\label{}
{\det}_N {\cal A}
:=\left|
\matrix{{\cal A}(q_1,q_1)& \cdots & {\cal A}(q_1,q_N)\cr
\vdots & \ddots & \vdots \cr
{\cal A}(q_N,q_1)& \cdots & {\cal A}(q_N,q_N)}
\right|.
\end{equation}
Functions ${\cal A}(q_a,q_b)$ are obtained from the elements
$A(q_a,q_b)$ as
\begin{equation} \label{lim}
{\cal A}(q_a,q_b) = \lim_{L\to\infty} \frac{L}{2\pi} A (q_a,q_b).
\end{equation}
The linear integral operator $\hat{\cal A}$ acts on functions $f(q)$ as
\begin{equation} \label{rule}
(\hat{\cal A} f)(q)=\int\limits_{-\infty}^{\infty} dq^{\prime}
{\cal A}(q,q^{\prime}) f(q^{\prime})
\end{equation}
and the kernel ${\cal A}(q,q')$ is given just by the equation
(\ref{lim}).

It should be kept in mind that in our notations $A(q_a,q_b)$ denote
matrix elements of matrices $A$ entering the expansions of the kind
(\ref{FA}) which contain the sums. Quantities ${\cal A}(q_a,q_b)$
denote matrix elements of matrices entering the expressions similar to
(\ref{Fredholm}) containing the integrals. Finally, $\hat{\cal A}$
denote the integral operator with the kernel ${\cal A}(q,q')$.

The Fredholm determinant  is well defined if the trace of operator
$\hat{\cal A}$ is finite $\int_{}^{}dq {\cal A}(q,q)<\infty$.  It is
the case for all the operators considered below with the exception of
the integral operator corresponding to the partition function which is
divergent in the limit (see (\ref{partition})). The partition function
is the denominator in the expressions (\ref{T-cor}) for the correlation
functions. The mean values in the numerators are also divergent. It
will be seen that this divergency is described by the same operator as
in the partition function, so that the final answer for the correlator
is finite. To make sense to the intermediate formulae, one should keep
this divergent operator regularized. We will make this regularization
writing for the partition function at $L\to\infty$
\begin{equation} \label{dom-fin}
\Tr\left[e^{-H/T}\right] = \det(\hat I+ \hat{\cal Z}), \qquad
{\cal Z}(q,q')
=2\cosh\!\matrix{\frac{B}{T}} e^{-\frac{q^2-h}{T}} \delta_{L}(q-q'),
\end{equation}
Here $\delta_{L}(q-q')$ is a regularization of the Dirac
delta-function; e.g., one can put
\begin{equation} %\label{}
\delta_{L}(q-q')\equiv\frac{\sin L(q-q')}{2\pi (q-q')}.
\end{equation}
Of course, this regularization reproduces only the leading term
of the partition function; the $1/L$ corrections to the free energy
should be calculated from the exact expression (\ref{form}).

Turn now to the correlator $G_1^{(-)}$ (\ref{T-cor}). Let us calculate
the numerator of $G_1^{(-)}$. There are three main steps in the
calculation: i) take the thermodynamic limit; ii) sum over $\lambda$'s
(or $\mu$'s) and $M$; iii) sum over $N$ by means of the Fredholm
determinant formula (\ref{Fredholm}), extracting explicitly the
determinant $\det(\hat I+ \hat{\cal Z})$ (which cancel exactly the
denominator). The result after taking $L\to\infty$ is the well defined
Fredholm determinant representation for the correlation function.

Consider the numerator of $G_1^{(-)}$. Using the representation
(\ref{NMV_1}) for the normalized mean value involved, we get
\begin{eqnarray} \label{numL}
\lefteqn{
\Tr\left[e^{-H/T} \psi^+_1(x,t)\psi^{}_1(0,0)\right]
}
\nonumber\\ &&
=\sum_{N=0}^{\infty}
\sum_{M=0}^{N+1}
\sum_{\tilde k_1<\ldots<\tilde k_{N+1}}^{}
\sum_{\lambda_1<\ldots<\lambda_M}^{} e^{-\frac{E_{N+1,M}(\{k\})}{T}}
\langle\psi^+_1(x,t) \psi^{}_1(0,0)\rangle_{N+1,M}
\nonumber\\ &&
=e^{-it(h-B)} \sum_{N=0}^{\infty}
\frac{1}{N}\sum_{p,n=0}^{N-1}
e^{-\frac{2\pi i}{N}pn}
\sum_{M=0}^{N+1} e^{\frac{2B}{T}M}
\sum_{\lambda_1<\ldots<\lambda_M}^{}
{\det}_M U^{(1,-)}_p
\nonumber\\ && \times
\sum_{\tilde k_1<\ldots<\tilde k_{N+1}}^{}
\left[{\det}_{N+1}(\widetilde S^{(-)}_n+\widetilde R^{(-)})
-{\det}_{N+1}\widetilde S^{(-)}_n\right],
\end{eqnarray}
where tildes over matrices mean that the temperature factors
$\exp\{-\sum_{a=1}^{N+1}(k_a^2-h+B)/T\}$ are included into the
determinants, i.e., the matrix elements $(\ )_{ab}$ of the corresponding
matrix is multiplied by the factor $\exp\{-\frac{k_a^2-h+B}{T}\}$.
It means that the matrix $\widetilde S^{(-)}_n$ has the
structure
\begin{equation} %\label{}
(\widetilde S^{(-)}_n)_{ab}
=e^{-\frac{k_a^2-h+B}{T}}\delta_{ab}+\cdots,\qquad
a,b=1,\ldots,N+1.
\end{equation}

Since in (\ref{numL}) the summation is over all values
$\tilde k_1<\ldots<\tilde k_{N+1}$, there is the periodicity
in $\Lambda$ (recall that $k_a=\tilde k_a + \Lambda/L$) and one can
insert the Kronecker symbol on $Z_{N+1}$ (\ref{KronLambda}).

Then in
the thermodynamic limit, $L\to\infty$, the numerator of $G^{(-)}_1$
acquire the form
\begin{eqnarray} \label{tdl-num}
\lefteqn{
\Tr\left[e^{-H/T} \psi^+_1(x,t)\psi^{}_1(0,0) \right]
}
\nonumber\\ &&
= e^{-it(h-B)} \sum_{N=0}^{\infty}
\frac{1}{N(N+1)}\sum_{r,m=0}^{N}\sum_{p,n=0}^{N-1}
e^{\frac{2\pi i}{N+1}rm-\frac{2\pi i}{N}pn}
\nonumber\\ && \times
X_{N,p,r}
\int\limits_{k_1<\ldots<k_{N+1}}^{} dk_1\cdots dk_{N+1}\,
\Xi^{(-)}_{N+1}(\eta_{n,m}),
\end{eqnarray}
where we denote
\begin{eqnarray} \label{XXi}
\lefteqn{
X_{N,p,r}:=
\sum_{M=0}^{N+1} e^{\frac{2B}{T}M}
\sum_{\lambda_1<\ldots<\lambda_M}^{}
{\det}_M U_{p,r}^{(1,-)},
}
\nonumber\\
\lefteqn{
\Xi^{(-)}_{N+1}(\eta_{n,m}):=
{\det}_{N+1}(\widetilde {\cal S}^{(-)}(\eta_{n,m})
+\widetilde {\!\cal R}{}^{(-)})
-{\det}_{N+1}\widetilde {\cal S}^{(-)}(\eta_{n,m}).
}
\end{eqnarray}
The matrices $(U_{p,r}^{(1,-)})$ have matrix elements
\begin{equation} %\label{}
(U_{p,r}^{(1,-)})_{ab}=e^{-ir\lambda_a}(U_p^{(1,-)})_{ab}
\end{equation}
and elements of matrices $(U_p^{(1,-)})$ are given by (\ref{Up-}).
The matrix ${\cal S}^{(-)}(\eta_{n,m})$ has the structure
\begin{equation} \label{tildeS}
\widetilde {\cal S}^{(-)}(\eta_{n,m})=
{\cal N}+\frac{1+\e\,\cos\eta_{n,m}}{2}\ {\widetilde {\cal V}_1}
+\frac{\e\,\sin\eta_{n,m}}{2}\ {\widetilde {\cal V}_2},
\end{equation}
where
\begin{equation} %\label{}
\eta_{n,m}:=\frac{2\pi m}{N+1}-\frac{2\pi n}{N}.
\end{equation}
The matrix elements of the matrix ${\cal N}$ are
\begin{equation} \label{N}
{\cal N}(k_a,k_b)=e^{-\frac{k_a^2-h+B}{T}}\,\delta_L(k_a-k_b)
\end{equation}
and the elements of the matrices $\widetilde {\cal V}_{1,2}$,
$\widetilde {\!\cal R}{}^{(-)}$
are obtained in accordance with the rule (\ref{rule}):
\begin{eqnarray} \label{tildeVQR}
\lefteqn{
\widetilde  {\cal V}_1(k_a,k_b)
=e^{-\frac{k_a^2-h+B}{T}}\
\frac{E_{+}(k_a)E_{-}(k_b)-E_{-}(k_a)E_{+}(k_b)}{\pi (k_a-k_b)},
}
\nonumber\\
\lefteqn{
\widetilde {\cal V}_2(k_a,k_b)
=e^{-\frac{k_a^2-h+B}{T}}\
\frac{[E_{-}(k_a)]^{-1}E_{-}(k_b)-E_{-}(k_a)[E_{-}(k_b)]^{-1}
}{\pi (k_a-k_b)},
}
\nonumber\\
\lefteqn{
\widetilde {\!\cal R}{}^{(-)}(k_a,k_b)
=e^{-\frac{k_a^2-h+B}{T}}\
\frac{E_{-}(k_a)E_{-}(k_b)}{2\pi}.
}
\end{eqnarray}
(Since the elements of the matrices $\widetilde {\cal V}^{(-)}_{1,2}$
and $\widetilde {\cal V}^{(+)}_{1,2}$, which are obtained from
$(\widetilde V_{1,2}^{(-)})_{ab}$ and $(\widetilde V_{1,2}^{(+)})_{ab}$
by means of (\ref{rule}), are described by the same functions,
there is no need to distinguish between
$\widetilde {\cal V}^{(-)}_{1,2}$ and
$\widetilde {\cal V}^{(+)}_{1,2}$, hence we write simply
$\widetilde {\cal V}_{1,2}$.)
The functions $E_{\pm}$ are
\begin{equation} \label{Epm}
E_{+}(k)= E(k)\ E_{-}(k), \qquad
E_{-}(k)=e^{\frac{itk^2-ixk}{2}},
\end{equation}
where
\begin{equation} \label{E(k)}
E(k)
=P.v.\int\limits_{-\infty}^{\infty}dq\,
\frac{e^{-itq^2+ixq}}{\pi (q-k)}.
\end{equation}
These functions are the thermodynamic limits of the corresponding
functions $e_{+}^{(\pm)}$, $e_{-}$, $e^{(\pm)}$ which is explained in
detail in Appendix.

Consider the quantity $X_{N,p,r}$. Note that, by definition,
$\left.{\det}_M U_{p,r}^{(1,-)}\right|_{M=0}=1$ and it is easy
to see that $\left.{\det}_M U_{p,r}^{(1,-)}\right|_{M=N+1}=0$.
Thus
\begin{equation} %\label{}
X_{N,p,r}= 1 + \sum_{M=1}^{N} e^{\frac{2B}{T}M}
\sum_{\lambda_1<\ldots<\lambda_M}^{} {\det}_M U_{p,r}^{(1,-)}.
\end{equation}

To obtain a useful expression for $X_{N,p,r}$ let us note that
the following relation is valid
\begin{equation} \label{DV=DW}
\sum_{\lambda_1<\ldots<\lambda_M}^{} {\det}_M U_{p,r}^{(1,-)}
=\sum_{1\leq m_1<\ldots<m_M\leq N}^{} {\det}_M W_{p,r}^{(1)},
\end{equation}
where
\begin{eqnarray} \label{detW}
\lefteqn{
{\det}_M W_{p,r}^{(1)}= \sum_{P\in S_M}^{}(-1)^{[P]}
}
\nonumber\\ && \times
\prod_{a=1}^{M}\left[\sum_{n_a=1}^{N} \biggl(\frac{1}{N+1}
\sum_{\lambda}^{}e^{-i\lambda (r+m_{P_a}-n_a)}\biggr)
\biggl(\frac{1}{N}\sum_{\mu}^{}e^{i\mu (p+m_a-n_a)}\biggr)\right].
\end{eqnarray}
To see that the relation (\ref{DV=DW})
holds it is sufficient to note that: i) ${\det}_M U_{p,r}^{(1,-)}$
is symmetric with respect to permutations of $\lambda$'s
being equal to zero whenever two of them coincide; ii) ${\det}_M
W_{p,r}^{(1)}$ is symmetric with
respect to permutations of $m$'s
(being zero whenever two of them coincide).
After changing $\sum_{\lambda_1<\ldots<\lambda_M}^{}$
to $\frac{1}{M!}\sum_{\lambda_1}^{}\cdots\sum_{\lambda_M}^{}$ in
the l.h.s. of (\ref{DV=DW})
(and similarly for the summation over $m$'s in the r.h.s.)
the relation (\ref{DV=DW}) becomes obvious.

Let us find the explicit expression for the elements of the matrix
$W_{p,r}^{(1)}$ .
Using Bethe equations for $\lambda$'s and $\mu$'s we have the following
summation formulae (here $m:=m_1,\ldots,m_M$; $n:=n_1,\ldots,n_M$):
\begin{eqnarray} \label{sm-sl}
\lefteqn{
\frac{1}{N+1} \sum_{\lambda}^{}e^{-i\lambda (r+m-n)}
}
\nonumber\\ &&
=\delta_{r+m-n}\theta (r+m<N+1)
+(-1)^{M+1}\delta_{r+m-n-N-1}\theta (r+m>N+1),
\nonumber\\
\lefteqn{
\frac{1}{N} \sum_{\mu}^{}e^{i\mu (p+m-n)}
}
\nonumber\\ &&
=\delta_{p+m-n} \theta (p+m\leq N)
+(-1)^{M+1}  \delta_{p+m-n-N} \theta (p+m>N).
\end{eqnarray}
These expressions are valid
only for the values of the integers involved: $p=0,\ldots,N-1$;
$r=0,\ldots,N$; $m_1,\ldots,m_M,n_1,\ldots,n_M=1,\ldots,N$. Here and
below we use the short notation for the usual Kronecker symbol,
$\delta_{a}=\delta_{a,0}$. Making use of (\ref{sm-sl}) and summing over
$n_1,\ldots,n_N$ in (\ref{detW}) we get
\begin{eqnarray} %\label{}
\lefteqn{
(W_{p,r}^{(1)})_{ab}=
\left[\delta_{p-r+m_a-m_b}+(-1)^{M+1}\delta_{p-r+m_a-m_b+N+1}\right]
\theta (p+m_a\leq N)
}
\nonumber\\ &&
+\left[\delta_{p-r+m_a-m_b+1}+(-1)^{M+1}\delta_{p-r+m_a-m_b-N}\right]
\theta (p+m_a>N).
\end{eqnarray}
Using the indentity for the determinant of an $N\times N$ matrix,
\begin{equation} %\label{}
{\det}_N (I+A) = 1 + \sum_{j=1}^{N} A_{j,j}
+ \sum_{j_1,j_2=1\atop j_1<j_2}^{N}
\left|
\matrix{
A_{j_1,j_1} & A_{j_1,j_2} \cr
A_{j_2,j_1} & A_{j_2,j_2}
}
\right| + \cdots + {\det}_N A,
\end{equation}
we obtain, due to the relation (\ref{DV=DW}), that
\begin{eqnarray} \label{4det}
\lefteqn{
X_{N,p,r}=
\frac{1}{2}\left[
{\det}_N(I+e^{\frac{2B}{T}}W_{-})+{\det}_N(I-e^{\frac{2B}{T}}W_{-})
\right]
}
\nonumber\\ &&
+\frac{1}{2}\left[
{\det}_N(I+e^{\frac{2B}{T}}W_{+})-{\det}_N(I-e^{\frac{2B}{T}}W_{+})
\right]
\end{eqnarray}
where $N\times N$ matrices $W_{\pm}$ are
\begin{eqnarray} \label{Wpm}
\lefteqn{
(W_{\pm})_{ab}=
\left[\delta_{p-r+a-b} \pm \delta_{p-r+a-b+N+1}\right]
\theta (p+a\leq N)
}
\nonumber\\ &&
+\left[\delta_{p-r+a-b+1} \pm \delta_{p-r+a-b-N}\right]
\theta (p+a>N).
\end{eqnarray}

To proceed further, let us look at the quantity
$\Xi^{(-)}_{N+1}(\eta_{n,m})$ as a function of $\eta_{n,m}$. Obviously,
\begin{equation} \label{Xi}
\Xi^{(-)}_{N+1}(\eta_{n,m})
= \sum_{s=-N}^{N}
e^{is\eta_{n,m}} \Phi^{(-)}_s,
\end{equation}
where $\Phi^{(-)}_s$ are some combinations of
the elements of the matrices
${\cal N}$, $\widetilde {\cal V}_{1,2}$,
$\widetilde {\!\cal R}{}^{(-)}$ which do not depend on $n,m$.
After summation over $n$ and $m$ in (\ref{tdl-num}) the
terms with $s=0,\ldots,N$ in (\ref{Xi}) will produce $\delta_{p-r}$
while
the terms with $s=-N,\ldots,-1$ will produce $\delta_{p-r+1}$. Hence we
should take into account only the terms in (\ref{4det}) for which
either the condition $r=p$
or the condition $r=p+1$ is satisfied. It follows from (\ref{4det}) and
(\ref{Wpm}) that
\begin{equation} %\label{}
\left. X_{N,p,r}\right|_{r=p}=(1+e^{2B/T})^{N-p},\qquad
\left. X_{N,p,r}\right|_{r=p+1}=(1+e^{2B/T})^p.
\end{equation}
Let us introduce the notation
\begin{equation} %\label{}
\gamma=1+ e^{2B/T}.
\end{equation}
Then we can write the answer
\begin{equation}
\label{Xpr} X_{N,p,r}=1+
(\gamma^{N-p}-1)\delta_{p-r}+(\gamma^p-1)\delta_{p-r+1}+
\mbox{unessential terms}
\end{equation}
where ``unessential terms'' are terms constructed from the
Kronecker symbols
containing in their arguments $p,r$ and which are different from
those written in (\ref{Xpr}) explicitly.
These ``unessential terms'' will not contribute after the
summation over $n,m,p,r$ in (\ref{tdl-num}).

Substituting (\ref{Xpr}) into (\ref{tdl-num}) we obtain the following
expression under the sum over $N$ and $(N+1)$-fold
integration over $k$'s (except the overall factor $\exp\{-it(h-B)\}$)
\begin{eqnarray} \label{Nterm}
\lefteqn{
\frac{1}{(N+1)N}\sum_{m=0}^{N}\sum_{n=0}^{N-1}
\Biggl\{\gamma^N-1
}
\nonumber\\  &&
+\sum_{p=1}^{N-1}(\gamma^{N-p}-1)
\Bigl[e^{2\pi i \eta_{n,m} p}+e^{-2\pi i \eta_{n,m} p}\Bigr]
\Biggr\}\Xi^{(-)}_{N+1}(\eta_{n,m})
+\Xi^{(-)}_{N+1}(0).
\end{eqnarray}

Due to the explicit dependence on $N$ of the elements of the matrix
under the determinant in (\ref{tdl-num}), some efforts are needed
in order to apply the Fredholm determinant formula (\ref{Fredholm})
to the expression
(\ref{tdl-num}) as a whole. The first trick is to
change the double sum (over $n,m$)
in the first term in (\ref{Nterm}) to the ordinary integral, i.e., the
following indentity holds
\begin{eqnarray} \label{basic}
\lefteqn{
\frac{1}{(N+1)N}\sum_{m=0}^{N}\sum_{n=0}^{N-1}
\Biggl\{ \gamma^N-1
}
\nonumber\\ &&
+\sum_{p=1}^{N-1}(\gamma^{N-p}-1)
\Bigl[e^{2\pi i \eta_{n,m} p}+e^{-2\pi i \eta_{n,m} p}\Bigr]
\Biggr\}\Xi^{(-)}_{N+1}(\eta_{n,m})
\nonumber\\ &&
=\frac{1}{2\pi}\int\limits_{-\pi}^{\pi} d\eta
\left\{ \gamma^N-1+\sum_{p=1}^{N-1}(\gamma^{N-p}-1)
\Bigl[e^{i\eta p}+ e^{-i\eta p}\Bigr] \right\} \Xi^{(-)}_{N+1}(\eta).
\end{eqnarray}
It is sufficient to sum over $n,m$ in the l.h.s. of (\ref{basic}) and
to integrate over $\eta$ in the r.h.s. in order to see that the results
coincide.

The second trick is to expand the summation with respect to $p$ in the
r.h.s. of (\ref{basic}) to the infinity. Adding the term at $p=N$
obviously
do not produce any additional terms to the r.h.s. of (\ref{basic})
because $ (\gamma^{N-p}-1)|_{p=N}=0$. The terms with $p>N$ will die
after taking the integration with respect to $\eta$ since then the
quantity $e^{i\eta p}\Xi_N(\eta)$ (or $e^{-i\eta p}\Xi_N(\eta)$)
contains $ e^{i\eta}$ only in integer positive (or only in
integer negative) powers, due to (\ref{Xi}).

Therefore, the expression (\ref{Nterm}) can be rewritten as
\begin{eqnarray} \label{Ntermfin}
\lefteqn{
\frac{1}{2\pi}\int\limits_{-\pi}^{\pi} d\eta
\left\{ \gamma^N-1+\sum_{p=1}^{\infty}(\gamma^{N-p}-1)
\Bigl[e^{i\eta p}+ e^{-i\eta p}\Bigr] \right\} \Xi^{(-)}_{N+1}(\eta)
+\Xi^{(-)}_{N+1}(0)
}
\nonumber\\ &&
=\frac{1}{2\pi}\int\limits_{-\pi}^{\pi} d\eta
\left\{ \gamma^N+\sum_{p=1}^{\infty}\gamma^{N-p}
\Bigl[e^{i\eta p}+ e^{-i\eta p}\Bigr] \right\} \Xi^{(-)}_{N+1}(\eta)
\nonumber\\ &&
=\frac{1}{2\pi}\int\limits_{-\pi}^{\pi} d\eta
\left\{1+\sum_{p=1}^{\infty}\gamma^{-p}
\Bigl[e^{i\eta p}+ e^{-i\eta p}\Bigr] \right\}
\Xi_{N+1}^{(-)}(\gamma;\eta).
\end{eqnarray}
Here we denote by $\Xi^{(-)}_{N+1}(\gamma;\eta)$ the expression for
$\Xi^{(-)}_{N+1}(\eta)$ given by (\ref{XXi}) where the replacement
$\widetilde {\cal S}^{(-)}(\eta)
\to\gamma\widetilde {\cal S}^{(-)}(\eta)$
has been
made.  The last expression in (\ref{Ntermfin}) is what we need in order
to apply the Fredholm formula (\ref{Fredholm}). Taking
into account that $\gamma{\cal N}(k_a,k_b)={\cal Z}(k_a,k_b)$, we get
\begin{eqnarray} %\label{}
\lefteqn{
\Tr\left[e^{-H/T} \psi^+_1(x,t)\psi^{}_1(0,0)\right]
=e^{-it(h-B)}\frac{1}{2\pi}\int\limits_{-\pi}^{\pi} d\eta\,
F(\gamma,\eta)
}
\nonumber\\ && \times
\biggl[
\det\left(\hat I+\hat {\cal Z}+\gamma \hat{\widetilde {\cal V}}(\eta)
+\hat{\widetilde {\!\cal R}}{}^{(-)}\right)
-\det\left(\hat I+\hat {\cal Z}+\gamma \hat{\widetilde {\cal V}}(\eta)
\right)\biggr].
\end{eqnarray}
Here
\begin{equation} \label{Veta}
\hat{\widetilde {\cal V}}(\eta):=
\frac{1+\e\,\cos\eta}{2}\ \hat{\widetilde {\cal V}}_1
+\frac{\sin\eta}{2}\ \hat{\widetilde {\cal V}}_2
\end{equation}
and $\hat {\cal Z}$, $\hat{\widetilde {\cal V}}_{1,2}$,
$\hat{\widetilde {\!\cal R}}{}^{(-)}$
are integral operators with the kernels given by (\ref{dom-fin}),
(\ref{tildeVQR}). We have introduced the function
\begin{equation} \label{Fgammaeta}
F(\gamma,\eta)=1+\sum_{p=1}^{\infty} \gamma^{-p}
(e^{i\eta p}+e^{-i\eta p}).
\end{equation}
It is worth mentioning that $1\leq \gamma <\infty$ for any real
external field $B$. At the point $\gamma=1$ ($B=-\infty$)
the function $F(\gamma,\eta)$ is equal (up to the factor $2\pi$) to the
$2\pi$-periodic delta-function of the variable $\eta$.

Finally, taking into account the representation (\ref{dom-fin}) for
$\Tr[e^{-H/T}]$ we obtain the following representation for the
temperature correlation function considered
\begin{eqnarray} \label{Gfin}
\lefteqn{
G^{(-)}_1(x,t;h,B)
=e^{-it(h-B)}\frac{1}{2\pi}\int\limits_{-\pi}^{\pi} d\eta\,
F(\gamma,\eta)
}
\nonumber\\ && \times
\biggl[\det\left(\hat I+\gamma\,\hat{\cal V}(\eta)+\hat{\cal R}^{(-)}
\right)
-\det\left(\hat I+\gamma\,\hat{\cal V}(\eta)\right)\biggr],
\end{eqnarray}
where integral operator $\hat{\cal V}(\eta)$ is of the form
\begin{equation} %\label{}
\hat{\cal V}(\eta)
=\frac{1+\e\,\cos\eta}{2}\ \hat{\cal V}_1
+\frac{\sin\eta}{2}\ \hat{\cal V}_2.
\end{equation}
The integral operators $\hat{{\cal V}}_{1,2}$, $\hat{{\cal R}}^{(-)}$
do not depend on $\eta$ and possess kernels
\begin{eqnarray} \label{calVQR}
\lefteqn{
{\cal V}_1(k,k')
=\frac{E^T_{+}(k) E^T_{-}(k')-E^T_{+}(k') E^T_{-}(k)}{\pi (k-k')},
}
\nonumber\\
\lefteqn{
{\cal V}_2(k,k')
=\frac{1}{\pi (k-k')}
\left[\frac{\vartheta(k)}{E^T_{-}(k)} E^T_{-}(k')
-E^T_{-}(k)\frac{\vartheta(k')}{E^T_{-}(k')}\right],
}
\nonumber\\
\lefteqn{
{\cal R}^{(-)}(k,k')
=\frac{E^T_{-}(k) E^T_{-}(k')}{2\pi}.
}
\end{eqnarray}
The functions $E^T_{+}(k)$, $E^T_{-}(k)$ are
\begin{equation} \label{ETs}
E^T_{+}(k) = E(k)\ E^T_{-}(k),
\qquad E^T_{-}(k)= \sqrt{\vartheta(k)}\, e^{\frac{itk^2-ixk}{2}},
\end{equation}
and the function $E(k)$ is given by (\ref{E(k)}). The Fermi weight
$\vartheta(k)$ is
\begin{equation} \label{Fermi}
\vartheta(k)=\frac{e^{-B/T}}{2\cosh\frac{B}{T}+e^{\frac{k^2-h}{T}}}.
\end{equation}

Let us now consider another temperature correlation function
$G^{(+)}_1$ (\ref{T-cor}).
Using the representation (\ref{NMV+})
for the normalized mean value, we represent the numerator of the
temperature correlation function as follows
\begin{eqnarray} \label{numLp}
\lefteqn{
\Tr\left[e^{-H/T} \psi^{}_1(x,t)\psi^+_1(0,0)\right]
}
\nonumber\\ &&
=\sum_{N=0}^{\infty} \sum_{M=0}^{N}
\sum_{\tilde q_1<\ldots<\tilde q_N}^{}
\sum_{\mu_1<\ldots<\mu_M}^{} e^{-\frac{E_{N,M}(\{q\})}{T}}
\langle\psi^{}_1(x,t) \psi^+_1(0,0)\rangle_{N,M}
\nonumber\\ &&
=e^{it(h-B)} \sum_{N=0}^{\infty}
\frac{1}{N+1}\sum_{r,m=0}^{N}
e^{\frac{2\pi i}{N+1}rm}
\sum_{M=0}^{N} e^{\frac{2B}{T}M}
\sum_{\mu_1<\ldots<\mu_M}^{}
{\det}_M U^{(1,+)}_r
\nonumber\\ && \times
\sum_{\tilde q_1<\ldots<\tilde q_N}^{}
\left[{\det}_N(\widetilde S^{(+)}_m-\widetilde R^{(+)}_m)
+(g_m(x,t)-1)\,{\det}_N\widetilde S^{(+)}_m\right],
\end{eqnarray}
where the tilde over the matrix means, as above, that the
temperature factor $ \exp\{-\sum_{a=1}^{N+1}(k_a^2-h+B)/T\}$ is
included into the determinant.

In the thermodynamic limit the numerator of $G^{(+)}_1$
acquire the form
\begin{eqnarray} \label{tdl-num2}
\lefteqn{
\Tr\left[e^{-H/T}\psi^{}_1(x,t)\psi^+_1(0,0)\right]
}
\nonumber\\ &&
= e^{it(h-B)} \sum_{N=0}^{\infty}
\frac{1}{N(N+1)}\sum_{r,m=0}^{N}\sum_{p,n=0}^{N-1}
e^{\frac{2\pi i}{N+1}rm-\frac{2\pi i}{N}pn}
\nonumber\\ && \times
X_{N,p,r}
\int\limits_{q_1<\ldots<q_N}^{} dq_1\cdots dq_N\
\Xi^{(+)}_{N}(\eta_{n,m})
\end{eqnarray}
where we denote
\begin{eqnarray} \label{XXi2}
\lefteqn{
\Xi^{(+)}_{N}(\eta_{n,m})
}
\nonumber\\ &&
:={\det}_N(\widetilde {\cal S}^{(+)}(\eta_{n,m})
-\widetilde {\!\cal R}{}^{(+)}(\eta_{n,m}))
+(G(x,t)-1)\,{\det}_N \widetilde {\cal S}^{(+)}(\eta_{n,m}).
\end{eqnarray}
The function $G(x,t)$ is the thermodynamic limit of the function
$g(x,t)$ (\ref{g(x,t)}):
\begin{equation} %\label{}
G(x,t)=\frac{1}{2\pi}\int\limits_{-\infty}^{\infty}dk e^{-itk^2+ixk}.
\end{equation}
The matrices $\widetilde {\cal S}^{(+)}(\eta_{n,m})$ and
$\widetilde {\!\cal R}{}^{(+)}(\eta_{n,m})$ have the structure
\begin{eqnarray} %\label{}
\lefteqn{
\widetilde {\cal S}^{(+)}(\eta_{n,m})=
{\cal N}+\frac{1+\e\,\cos\eta_{n,m}}{2}\ {\widetilde {\cal V}_1}
-\frac{\e\,\sin\eta_{n,m}}{2}\ {\widetilde {\cal V}_2},
}
\nonumber\\
\lefteqn{
\widetilde {\!\cal R}{}^{(+)}(\eta_{n,m})=
\frac{1+\e\,\cos\eta_{n,m}}{2}\ \widetilde {\!\cal R}{}^{(+)}_1
}
\nonumber\\ &&
-\e\,\sin\eta_{n,m}\ \widetilde {\!\cal R}{}^{(+)}_2 +
\frac{1-\e\,\cos\eta_{n,m}}{2}\ \widetilde {\!\cal R}{}^{(+)}_3.
\end{eqnarray}
Matrix elements of the matrices ${\cal N}$ and
$\widetilde {\cal V}_{1,2}$ (which are $N\times N$ dimensional
matrices here) are given
by the equations (\ref{N}) and (\ref{tildeVQR})
(but with the momenta $q$ instead of $k$).
The elements of the matrices
$\widetilde {\!\cal R}{}^{(+)}_{1,2,3}$ (functions
$\widetilde {\!\cal R}{}^{(+)}_{1,2,3}(q_a,q_b)$) are
\begin{eqnarray} \label{tildeR123}
\lefteqn{
\widetilde {\!\cal R}{}^{(+)}_1(q_a,q_b)
=e^{-\frac{q_a^2-h+B}{T}}\,
\frac{E_{+}(q_a) E_{+}(q_b)}{2\pi},
}
\nonumber\\
\lefteqn{
\widetilde {\!\cal R}{}^{(+)}_2(q_a,q_b)
=e^{-\frac{q_a^2-h+B}{T}}\,
\frac{1}{4\pi}\left[
\frac{E_{+}(q_a)}{E_{-}(q_b)}+\frac{E_{+}(q_b)}{E_{-}(q_a)}
\right],
}
\nonumber\\
\lefteqn{
\widetilde {\!\cal R}{}^{(+)}_3(q_a,q_b)
=e^{-\frac{q_a^2-h+B}{T}}\,
\frac{1}{2\pi E_{-}(q_a) E_{-}(q_b)}.
}
\end{eqnarray}
The functions $E_{\pm}$, entering
(\ref{tildeR123}) are defined in (\ref{Epm}), (\ref{E(k)}).

In (\ref{tdl-num2}), the quantity $X_{N,p,r}$
is the same as in (\ref{XXi}) above, since
\begin{eqnarray} %\label{}
\lefteqn{
X_{N,p,r}
:=\sum_{M=0}^{N}e^{\frac{2B}{T}M}
\sum_{\mu_1<\ldots<\mu_M}^{}
{\det}_M U_{p,r}^{(1,+)}(\{\mu\})
}
\nonumber\\ &&
=\sum_{M=0}^{N+1} e^{\frac{2B}{T}M}
\sum_{\lambda_1<\ldots<\lambda_M}^{}
{\det}_M U_{p,r}^{(1,-)}(\{\lambda\}),
\end{eqnarray}
where
\begin{equation}
(U_{p,r}^{(1,+)}(\{\mu\}))_{ab}=e^{ip\mu_a}(U_r^{(1,+)}(\{\mu\}))_{ab}
\end{equation}
and the elements of the matrices $U_r^{(1,+)}$ are given by
(\ref{Up+}).

Similarly to (\ref{Xi}), we have
\begin{equation} \label{Xi2}
\Xi^{(+)}_{N}(\eta_{n,m}) = \sum_{s=-N}^{N}
e^{is\eta_{n,m}} \Phi^{(+)}_s,
\end{equation}
where $\Phi^{(+)}_s$ are some functions which do not depend on $n,m$.
Therefore, the result (\ref{Xpr}) for $X_{N,p,r}$ is valid also in the
case considered and we can repeat the calculation described by
equations (\ref{Nterm})--(\ref{Ntermfin}),
except that we should write
$\Xi_N^{(+)}$ instead of $\Xi_{N+1}^{(-)}$.
The last expression in (\ref{Ntermfin}) is now written as
\begin{equation} %\label{}
\frac{1}{2\pi}\int\limits_{-\pi}^{\pi} d\eta
\left\{1+\sum_{p=}^{\infty}\gamma^{-p}
\Bigl[e^{i\eta p}+ e^{-i\eta p}\Bigr]
\right\} \Xi_N^{(+)}(\eta;\gamma).
\end{equation}
Here $\Xi^{(+)}_N(\eta;\gamma)$ denotes the expression for
$\Xi^{(+)}_N(\eta)$ given by (\ref{XXi2})
in which the replacement
$\widetilde {\cal S}^{(+)}(\eta)
\to\gamma\widetilde {\cal S}^{(+)}(\eta)$,
$\widetilde {\!\cal R}{}^{(+)}(\eta)
\to\gamma\widetilde {\!\cal R}{}^{(+)}(\eta)$ has been made.

Thus we obtain the Fredholm determinant representation
for the numerator of the correlator $G^{(+)}_1$
\begin{eqnarray} %\label{}
\lefteqn{
\Tr\left[e^{-H/T} \psi^{}_1(x,t)\psi^+_1(0,0)\right]
}
\nonumber\\ &&
=e^{it(h-B)}\frac{1}{2\pi}\int\limits_{-\pi}^{\pi} d\eta\,
F(\gamma,\eta)
\biggl[\det\left(\hat I+\hat {\cal Z}
+\gamma\,\hat{\widetilde {\cal V}}(\eta)
-\gamma\,\hat{\widetilde {\!\cal R}}{}^{(+)}(\eta)\right)
\nonumber\\ &&
+(G(x,t)-1)\det\left(\hat I+\hat {\cal Z}
+\gamma\,\hat{\widetilde {\cal V}}(\eta)\right)\biggr],
\end{eqnarray}
where $\hat{\widetilde {\cal V}}(\eta)$ is given by (\ref{Veta})
and
\begin{equation}
\hat{\widetilde {\!\cal R}}{}^{(+)}(\eta):=
\frac{1+\e\,\cos\eta}{2}\ \hat{\widetilde {\!\cal R}}{}^{(+)}_1
+\sin\eta\ \hat{\widetilde {\!\cal R}}{}^{(+)}_2
+\frac{1-\e\,\cos\eta}{2}\ \hat{\widetilde {\!\cal R}}{}^{(+)}_3.
\end{equation}
The integral operators $\hat {\cal Z}$,
$\hat{\widetilde {\!\cal R}}{}^{(+)}_{1,2,3}$ possess
kernels given by the functions (\ref{dom-fin}), (\ref{tildeVQR}),
(\ref{tildeR123}) and the function $F(\gamma,\eta)$ is given by
(\ref{Fgammaeta}).

Finally, taking into account the representation (\ref{dom-fin}) for
$\Tr[e^{-H/T}]$ we obtain the following representation for the
temperature correlation function
\begin{eqnarray} \label{Gfin2}
\lefteqn{
G^{(+)}_1(x,t;h,B)
=e^{it(h-B)}\frac{1}{2\pi}\int\limits_{-\pi}^{\pi}d\eta\,F(\gamma,\eta)
}
\nonumber\\ && \times
\biggl[\det\left(\hat I+\gamma\,\hat{\cal V}(\eta)
-\gamma\,\hat{\cal R}^{(+)}(\eta)\right)
+(G(x,t)-1)\det\left(\hat I+\gamma\,\hat{\cal V}(\eta)\right)\biggr]
\end{eqnarray}
where
\begin{eqnarray} %\label{}
\lefteqn{
\hat{\cal V}(\eta)
=\frac{1+\e\,\cos\eta}{2}\ \hat{\cal V}_1
+\frac{\sin\eta}{2}\ \hat{\cal V}_2,
}
\nonumber\\
\lefteqn{
\hat{\cal R}^{(+)}(\eta)=
\frac{1+\e\,\cos\eta}{2}\ \hat{\cal R}^{(+)}_1
+\sin\eta\ \hat{\cal R}^{(+)}_2
+\frac{1-\e\,\cos\eta}{2}\ \hat{\cal R}^{(+)}_3.
}
\end{eqnarray}
The integral operators $\hat{{\cal V}}_{1,2}$ possess kernels
(\ref{calVQR}) and integral operators
$\hat{{\cal R}}^{(+)}_{1,2,3}$ possess kernels
\begin{eqnarray} \label{calR123}
\lefteqn{
{\cal R}^{(+)}_1(q,q')
=\frac{E^T_{+}(q) E^T_{+}(q')}{2\pi},
}
\nonumber\\
\lefteqn{
{\cal R}^{(+)}_2(q,q')
=\frac{1}{4\pi}\left[
E^T_{+}(q)\frac{\vartheta(q')}{E^T_{-}(q')}
+\frac{\vartheta(q)}{E^T_{-}(q)} E^T_{+}(q')
\right],
}
\nonumber\\
\lefteqn{
{\cal R}^{(+)}_3(q,q')
=\frac{1}{2\pi}
\frac{\vartheta(q)}{E^T_{-}(q)} \frac{\vartheta(q')}{E^T_{-}(q')}.
}
\end{eqnarray}
The functions $E^T_{+}$, $E^T_{-}$ are given by equations (\ref{ETs})
with the Fermi weight (\ref{Fermi}).

Let us now discuss our main results, the representations
(\ref{Gfin}) and (\ref{Gfin2}). The integral operators
$\hat{\cal V}(\eta)$ and $\hat{\cal R}^{(+)}(\eta)$
involved into these representations can be put in the form
usual to the integrable models, i.e.,
they are ``integrable integral operators''
in the sense of paper \cite{IIKS-93} (see also \cite{KBI}).
To show this we introduce a pair of functions
\begin{equation} %\label{}
\ell_{+}^T(\eta|k) :=
\frac{1+\e\,\cos\eta}{2}\ E_{+}^T(k)
+\frac{\sin\eta}{2}\ \frac{\vartheta(k)}{E_{-}^T(k)},\qquad
\ell_{-}^T(k) := E_{-}^T(k).
\end{equation}
The kernels of the operators can be put into the form
\begin{eqnarray} %\label{}
\lefteqn{
{\cal V}(\eta|k,k')
=\frac{\ell_{+}^T(\eta|k)\ \ell_{-}^T(k')-\ell_{-}^T(k)\
\ell_{+}^T(\eta|k')}
{\pi(k-k')},
}
\nonumber\\
\lefteqn{
{\cal R}^{(-)}(k,k')
=\frac{\ell_{-}^T(k)\ \ell_{-}^T(k')}{2\pi}, \qquad
{\cal R}^{(+)}(\eta|k,k')
=\frac{\ell_{+}^T(\eta|k)\ \ell_{+}^T(\eta|k')}{\pi(1+\e\,\cos\eta)}.
}
\end{eqnarray}
This form is important for deriving integrable partial differential
equations for the correlators and for constructing the corresponding
matrix Riemann-Hilbert problem.  This in turn will make possible the
evaluation of different asymptotics of the correlators considered.

%%%%%%%%%%%%%%%%%%%%%%%%%%%%%%%%%%%%%%%%%%%%%%%%%%%%%%%%%%%%%%%%%%%%%%%
\section{Particular cases}

Let us discuss now some particular cases of the representations
obtained.

First, consider the ``one-component'' limit of the theory.  Let
$B\to -\infty$, $h\to -\infty$ in such a way that $h_1=h-B$ is fixed.
Note that $h_2=h+B\to -\infty$ in this limit. It means that the
energy of the particles of type 2 becomes (see (\ref{E})) infinite and
therefore these particles are excluded from the spectrum of the theory.
So one should has the one-component gas of particles of type 1. Indeed,
the parameter $\gamma =1$ in this limit. Remind that the function
$F(\gamma =1,\eta)$ is proportional to the $2\pi$-periodic
delta-function of the variable $\eta$, and the integrals over $\eta$ in
(\ref{Gfin}), (\ref{Gfin2})
can be easily taken.  For $\e =+1$ one gets in
this way exactly the representations for the corresponding temperature
correlation functions of the impenetrable one-component Bose gas
obtained in papers \cite{KS-90,KS-91}:
\begin{eqnarray}%\label{}
\lefteqn{
\left. G_1^{(-)}(x,t;h,B)\right|_{h,B\to -\infty \atop h-B=h_1}
= e^{-ith_1} \left[\det(\hat I+\hat{{\cal V}}_1+
\hat{{\cal R}}^{(-)})- \det(\hat I+\hat{{\cal V}}_1)\right],
}
\nonumber\\
\lefteqn{
\left. G_1^{(+)}(x,t;h,B)\right|_{h,B\to -\infty \atop h-B=h_1}
}
\nonumber\\ &&
= e^{ith_1} \left[\det(\hat I+\hat{{\cal V}}_1
-\hat{{\cal R}}_1^{(+)})
+(G(x,t)-1)\det(\hat I+\hat{{\cal V}}_1)\right],
\end{eqnarray}
where the kernels of the integral operators $\hat{{\cal V}}_1$,
$\hat{{\cal R}}^{(-)}$, $\hat{{\cal R}}^{(+)}_1$
are given by (\ref{calVQR}) and (\ref{calR123})
with the Fermi weight
\begin{equation} %\label{}
\vartheta(k) =\frac{1}{1+e^{\frac{k^2-h_1}{T}}}.
\end{equation}
If $\e=-1$ (Fermi statistics)  then the correlators simplify
strongly, reproducing the well known simple results for the
one-component Fermi gas. The correlator
$G_2^{(-)}(x,t;h,B)=G_1^{(-)}(x,t;h,-B)$ vanish in the limit, what
means that  the particles of type 2 are really excluded from the
theory.

Second, consider the particular case of the equal time ($t=0$)
temperature correlators of the two-component gas. It appears that in
this case the integrals over $\eta$ in (\ref{Gfin}) and (\ref{Gfin2})
can be also taken with the result
\begin{eqnarray} \label{t=0}
\lefteqn{
G_1^{(-)}(x,0;h,B)
=\det\left(\hat I
+\frac{\gamma+\e}{2}\ \hat v + \hat r^{(-)}\right)
-\det\left(\hat I+\frac{\gamma+\e}{2}\ \hat v\right),
}
\nonumber\\
\lefteqn{
G_1^{(+)}(x,0;h,B)
}
\nonumber\\ &&
=\det\left(\hat I
+\frac{\gamma+\e}{2}\ \hat v
+\e\,\hat r^{(+)}\right)
+(\delta(x)-1)
\det\left(\hat I +\frac{\gamma+\e}{2}\ \hat v\right),
\end{eqnarray}
where the integral operators $\hat v$, $\hat r^{(\pm)}$ possess kernels
\begin{eqnarray} %\label{}
\lefteqn{
\hat v (k,k')=-\sqrt{\vartheta(k)}\,
\frac{2\sin (|x| \frac{k-k'}{2})}{\pi(k-k')}\,
\sqrt{\vartheta(k')},
}
\nonumber\\
\lefteqn{
\hat r^{(\pm)} (k,k')=\sqrt{\vartheta(k)}\,
\frac{\exp(\pm ix\frac{k+k'}{2})}{2\pi}\,\sqrt{\vartheta(k')},
}
\end{eqnarray}
with the Fermi weight $\vartheta(k)$ given by (\ref{Fermi}).
The equal-time correlators (\ref{t=0}) satisfy the relation
\begin{equation} %\label{}
G^{(+)}_1(x,0;h,B)=\delta(x)+\e\, G^{(-)}_1(-x,0;h,B),
\end{equation}
which could be also easily derived from the canonical commutation
relations (\ref{ecom}).  The representation (\ref{t=0}) for the
equal-time correlation function for the impenetrable Fermi gas
($\e=-1$) written in slightly different form has been obtained
in \cite{Berkovich}.

Discuss now the case of the two-component gas at zero temperature,
$T=0$. In this case one has
\begin{equation} %\label{}
\gamma\,\vartheta (k) = \theta(k_F^2-k^2),
\end{equation}
where $\theta$ is the step function and the Fermi momentum is
$k_F=\sqrt{h+|B|}$.

Consider first zero magnetic field $B=0$. In this case the ground state
is degenerate (see Section 1), the parameter $\gamma=2$, and one
obtains
\begin{eqnarray} %\label{}
\lefteqn{
\left. G^{(-)}_1(x,t;h,0)\right|_{T=0}
=\left. G^{(-)}_2(x,t;h,0)\right|_{T=0}
}
\nonumber\\ &&
=e^{-ith}\frac{1}{2\pi}\int\limits_{-\pi}^{\pi} d\eta\,
\frac{3}{5-4\cos\eta}
\nonumber\\ && \times
\biggl[\det\left(\hat I+\hat{\cal V}_0(\eta)
+\!\matrix{\frac{1}{2}}\hat{\cal R}_0^{(-)}\right)
-\det\left(\hat I+\hat{\cal V}_0(\eta)\right) \biggr],
\nonumber\\
\lefteqn{
\left. G^{(+)}_1(x,t;h,0)\right|_{T=0}
=\left. G^{(+)}_2(x,t;h,0)\right|_{T=0}
}
\nonumber\\ &&
=e^{ith}\frac{1}{2\pi}\int\limits_{-\pi}^{\pi} d\eta\,
\frac{3}{5-4\cos\eta}
\nonumber\\ && \times
\biggl[\det\left(\hat I+\hat{\cal V}_0(\eta)
-\hat{\cal R}_0^{(+)}(\eta)\right)
+(G(x,t)-1)\det\left(\hat I+\hat{\cal V}_0(\eta)\right)\biggr].
\end{eqnarray}
Here the integral operators act on the
functions on the interval $(-k_F,k_F)$, e.g.,
\begin{equation} \label{FZ}
\left(\hat{\cal V}_0(\eta)\cdot f\right)(k)
=\int\limits_{-k_F}^{k_F}dk'\,
{\cal V}_0(\eta|k,k') f(k').
\end{equation}
The kernels are given as
\begin{eqnarray} \label{T=0}
\lefteqn{
{\cal V}_0(\eta|k,k')
=\frac{\ell_{+}(\eta|k)\ \ell_{-}(k')-\ell_{-}(k)\
\ell_{+}(\eta|k')}
{\pi(k-k')},
}
\nonumber\\
\lefteqn{
{\cal R}^{(-)}_0(k,k')
=\frac{\ell_{-}(k)\ \ell_{-}(k')}{2\pi}, \qquad
{\cal R}^{(+)}_0(\eta|k,k')
=\frac{\ell_{+}(\eta|k)\ \ell_{+}(\eta|k')}{\pi(1+\e\,\cos\eta)},
}
\end{eqnarray}
where
\begin{equation} %\label{}
\ell_{+}(\eta|k) :=
\frac{1+\e\,\cos\eta}{2}\ E_{+}(k)
+\frac{\sin\eta}{2}\ \frac{1}{E_{-}(k)},\qquad
\ell_{-}(k) := E_{-}(k),
\end{equation}
and functions $E_{\pm}(k)$ are given by (\ref{Epm}).

Turn to the case $B<0$. In this case the ground state is the
Fermi zone filled by the particles of type 1; one has $\gamma=1$ and
therefore $F(1,\eta)=2\pi \Delta(\eta)$ where $\Delta(\eta)$ is a
$2\pi$-periodic delta-function. For the correlators one obtains
\begin{eqnarray} %\label{}
\lefteqn{
\left. G_1^{(-)}(x,t;h,B)\right|_{B<0, T=0}
}
\nonumber\\ &&
= e^{-it(h-B)}
\biggl[\det\left(\hat I+\hat{{\cal V}}_0(0)+
\hat{{\cal R}}_0^{(-)}\right)
-\det\left(\hat I+\hat{{\cal V}}_0(0)\right)\biggr],
\nonumber\\
\lefteqn{
\left. G_1^{(+)}(x,t;h,B)\right|_{B<0, T=0}
}
\nonumber\\ &&
= e^{it(h-B)} \biggl[\det\left(\hat I+\hat{{\cal V}}_0(0)
-\hat{{\cal R}}_0^{(+)}(0)\right)
+(G(x,t)-1)\det\left(\hat I+\hat{{\cal V}}_0(0)\right)\biggr],
\nonumber\\
\lefteqn{
\left. G_2^{(-)}(x,t;h,B)\right|_{B<0, T=0} =0,
}
\nonumber\\
\lefteqn{
\left. G_2^{(+)}(x,t;h,B)\right|_{B<0, T=0}
=e^{it(h+B)}\frac{1}{2\pi}\int\limits_{-\pi}^{\pi} d\eta\,
}
\nonumber\\ && \times
\biggl[\det\left(\hat I+\hat{\cal V}_0(\eta)
-\hat{\cal R}_0^{(+)}(\eta)\right)
+(G(x,t)-1)\det\left(\hat I+\hat{\cal V}_0(\eta)\right)\biggr],
\end{eqnarray}
where
\begin{equation} %\label{}
\hat{\cal V}_0(0) = \frac{1+\e}{2} \hat{\cal V}_{1,0},\qquad
\hat{\cal R}^{(+)}_0(0)
=\frac{1+\e}{2} \hat{\cal R}^{(+)}_{1,0}
+\frac{1-\e}{2} \hat{\cal R}^{(+)}_{3,0}.
\end{equation}
The integral operators act on
the functions on the interval $(-k_F, k_F)$
by the rule (\ref{FZ}). The kernels are given by
\begin{eqnarray} %\label{}
\lefteqn{
{\cal V}_{1,0}(k,k')=
\frac{E_{+}(k) E_{-}(k')-E_{-}(k) E_{+}(k')}{\pi(k-k')},
}
\nonumber\\
\lefteqn{
{\cal R}_{1,0}^{(+)}(k,k')= \frac{E_{+}(k) E_{+}(k')}{2\pi},\qquad
{\cal R}_{3,0}^{(+)}(k,k')= \frac{1}{2\pi E_{-}(k) E_{-}(k')},
}
\end{eqnarray}
and by (\ref{T=0}).

If $B>0$, the ground state is formed by the particles of type 2, and
the formulae for the correlators are obtained in an obvious way,
due to the relation (\ref{2-1}).

%%%%%%%%%%%%%%%%%%%%%%%%%%%%%%%%%%%%%%%%%%%%%%%%%%%%%%%%%%%%%%%%%%%%%%%
\section*{Conclusion}

In this paper the determinant representation for the two-point
time-dependent temperature correlation functions of the one-dimensional
two-component bosons and fermions are obtained. The Fredholm
determinant of the integrable integral operator enters our answers
(\ref{Gfin}), (\ref{Gfin2}). It makes possible deriving intergrable
differential equations for the correlators and calculating their
large time and distance asymptotics.

From the technical point of view, the essential thing in our approach
is using the $XX0$ eigenfunctions for the two-component model with the
infinitely strong coupling. This gives the opportunity of explicit
calculation of the correlators. The same technique can
be applied for calculating temperature
correlation functions in the Hubbard model
in the infinitely strong coupling limit. We are going to present
the corresponding results in the next publication (with
N.\,I.\,Abarenkova).

%%%%%%%%%%%%%%%%%%%%%%%%%%%%%%%%%%%%%%%%%%%%%%%%%%%%%%%%%%%%%%%%%%%%%%%
\section*{Acknowlegments}

This work is supported in part by the grants RFFI No. 95-01-00476 and
INTAS-RFBR 95-0414.

%%%%%%%%%%%%%%%%%%%%%%%%%%%%%%%%%%%%%%%%%%%%%%%%%%%%%%%%%%%%%%%%%%%%%%%
\setcounter{section}{1} \def\thesection {\Alph{section}}

\section*{Appendix}

Here we consider matrix elements of the matrices
$S^{(-)}$, $S^{(+)}$, $R^{(+)}$ entering the representations for
normalized mean values for the model on the finite interval (Section 3).
Our aim is to rewrite matrix elements of these matrices in terms of
functions which are well defined in the thermodynamic limit.
We prove also the representations (\ref{IVQ}), (\ref{S+R+})
for the matrices considered.

We begin from a pair of indentities
\begin{equation} \label{cotsin}
\cot \pi z = \frac{1}{z} + \sum_{j=-\infty\atop j\ne 0}^{\infty}
\left(\frac{1}{z-\pi j} -\frac{1}{\pi j}\right),\qquad
\frac{\pi^2}{\sin^2 \pi z}= \sum_{j=-\infty}^{\infty}
\frac{1}{(z-\pi j)^2}.
\end{equation}
The value of the r.h.s. do not depend on organizing of summation.
Let us fix the method of summation in such a way that the
term $1/\pi j$ does not contribute to the sum in the first equation,
e.g., combining terms with the same $|j|$ together and summing up
in increasing order of $|j|=0,1,2,\ldots$.
The sums below should be understood with this
prescription (where necessary).

In terms of quasimomenta $k$ and $q$,
\begin{eqnarray} %\label{}
\lefteqn{
(k)_j = \frac{2\pi}{L}
\left(-\Bigl(\frac{1+\e}{2}\Bigr)\frac{N}{2}+j\right)+
\frac{\Lambda}{L},
}
\nonumber\\
\lefteqn{
(q)_j = \frac{2\pi}{L}
\left(-\Bigl(\frac{1+\e}{2}\Bigr)\frac{N-1}{2}+j\right)+
\frac{\Theta}{L},\qquad j\in Z,
}
\end{eqnarray}
the both indentities (\ref{cotsin}) become ($q=\tilde q+\Lambda/L$):
\begin{equation} \label{tan}
\frac{2}{L}\sum_{\tilde q}^{} \frac{1}{q-k}
=\e \left[\tan\biggl(\frac{\Lambda-\Theta}{2}\biggr)\right]^\e,\qquad
\frac{|1+\e\,\bar\omega\nu|^2}{L^2}
\sum_{\tilde q}^{} \frac{1}{(q-k)^2} = 1,
\end{equation}
where $\omega:=\exp i\Lambda$ and $\nu:=\exp i\Theta$.
The expression below, due to the first indentity in
(\ref{tan}), can be rewritten as
\begin{eqnarray} \label{s=ee}
\lefteqn{
\frac{|1+\e\bar\omega\nu|^2}{L} \sum_{\tilde q}^{}
\frac{e^{-itq^2+ixq}}{q-k}
}
\nonumber\\ &&
=\frac{|1+\e\bar\omega\nu|^2}{L}
\left\{\sum_{\tilde q}^{}
\frac{e^{-itq^2+ixq}-e^{-itk^2+ixk}}{q-k}
+\sum_{\tilde q}^{}
\frac{e^{-itk^2+ixk}}{q-k}
\right\}
\nonumber\\ &&
=\frac{1+\e\,\cos(\Lambda-\Theta)}{2} e^{(-)}(k)
+\frac{\e\,\sin(\Lambda-\Theta)}{2} [e_{-}(k)]^{-2}
\end{eqnarray}
where the functions
$e_{-}(k)$ and $e^{(-)}(k)$ are introduced,
\begin{eqnarray} \label{e^{(-)}(k)}
\lefteqn{
e^{(-)}(k)=\frac{2}{L}\sum_{\tilde q}^{}
\frac{e^{-itq^2+ixq}-e^{-itk^2+ixk}}{q-k},\qquad
q=\tilde q+\frac{\Theta}{L},
}
\nonumber\\
\lefteqn{
e_{-}(k)=\exp\left(\frac{itk^2-ixk}{2}\right).
}
\end{eqnarray}

Let us consider the elements of the matrix $S^{(-)}$ given by
\begin{equation} %\label{}
(S^{(-)})_{ab}
=e_{-}(k_a)\ e_{-}(k_b)\
\frac{|1+\e\nu\bar\omega|^{2}}{L^2}\
\sum_{\tilde q}^{} \frac{e^{-itq^2+ixq}}{(q-k_a)(q-k_b)}.
\end{equation}
Consider first the off-diagonal elements, i.e., the case when $a\ne b$,
hence $k_a \ne k_b$. For the off-diagonal elements of
matrix $S^{(-)}$, due to (\ref{s=ee}), one has
\begin{eqnarray} \label{off-diag}
\lefteqn{
(S^{(-)})_{ab}=e_{-}(k_a)\ e_{-}(k_b)\
\frac{|1+\e\nu\bar\omega|^2}{L^2(k_a-k_b)}
\sum_{\tilde q}^{}  \left[
\frac{e^{-itq^2+ixq}}{q-k_a}-\frac{e^{-itq^2+ixq}}{q-k_b} \right]
}
\nonumber\\ &&
=\frac{2}{L}\biggl\{
\frac{1+\e\,\cos(\Lambda-\Theta)}{2}\,
\frac{e^{(-)}_{+}(k_a) e_{-}(k_b)
-e_{-}(k_a) e^{(-)}_{+}(k_b)}{k_a-k_b}
\nonumber\\ &&
+\frac{\e\,\sin(\Lambda-\Theta)}{2}
\frac{[e_{-}(k_a)]^{-1} e_{-}(k_b)
-e_{-}(k_a) [e_{-}(k_b)]^{-1}}{k_a-k_b}
\biggr\}.
\end{eqnarray}
Let us consider now the diagonal elements
of the matrix $S^{(-)}$.
Using first the second indentity in (\ref{tan}) and then
(\ref{s=ee}), one has
for the diagonal elements
\begin{eqnarray} \label{diag}
\lefteqn{
(S^{(-)})_{aa}=
e^{itk_a^2-ixk_a}\left\{
\frac{|1+\e\bar\omega\nu|^2}{L^2}
\sum_{\tilde q}^{}
\frac{e^{-itq^2+ixq}-e^{-itk_a^2+ixk_a}}{(q-k_a)^2}
+e^{-itk_a^2+ixk_a}
\right\}
}
\nonumber\\ &&
=\frac{2}{L}
\Biggl\{\frac{1+\e\,\cos(\Lambda-\Theta)}{2}\,
\nonumber\\ && \times
e^{itk_a^2-ixk_a} \frac{2}{L}
\sum_{\tilde q}^{} \frac{e^{-itq^2+ixq}-e^{-itk_a^2+ixk_a}
\Bigl(1+i(x-2tk_a)(q-k_a)\Bigr)}{(q-k_a)^2}
\nonumber\\ &&
+ \frac{\e\,\sin(\Lambda-\Theta)}{2}\,
i(x-2tk_a)\Biggr\} + 1.
\end{eqnarray}
It is easy to see that the expressions in (\ref{diag}) following
$1+\e\cos(\Lambda-\Theta)$ and
$\e\sin(\Lambda-\Theta)$ are the limiting cases of
the corresponding expressions in (\ref{off-diag}) when $k_b\to k_a$.
Thus, collecting (\ref{off-diag}) together with (\ref{diag}), one
can conclude that the matrix $S^{(-)}$ has the structure (\ref{IVQ})
where matrices $V^{(-)}_{1,2}$ are given by (\ref{VQ}) with the
diagonal elements being understood in the sense of the l'H\^opitale
rule.

Consider now matrix elements of the matrices $S^{(+)}$ and $R^{(+)}$
given by (\ref{SR}). In this case, instead of (\ref{tan}), the
following indentities are useful
\begin{equation} \label{tan2}
\frac{2}{L}\sum_{\tilde k}^{} \frac{1}{k-q}
=-\e\left[\tan\biggl(\frac{\Lambda-\Theta}{2}\biggr)\right]^\e,\qquad
\frac{|1+\e\,\bar\omega\nu|^2}{L^2}
\sum_{\tilde k}^{} \frac{1}{(k-q)^2} = 1,
\end{equation}
which are, of course, direct consequences of (\ref{cotsin}).
Also, instead of (\ref{s=ee}), one has
\begin{eqnarray} \label{s=ee2}
\lefteqn{
\frac{|1+\e\bar\omega\nu|^2}{L} \sum_{\tilde k}^{}
\frac{e^{-itk^2+ixk}}{k-q}
}
\nonumber\\ &&
=\frac{1+\e\,\cos(\Lambda-\Theta)}{2} e^{(+)}(q)
-\frac{\e\,\sin(\Lambda-\Theta)}{2} [e_{-}(q)]^{-2}
\end{eqnarray}
where the function $e^{(+)}(q)$ is defined as
\begin{equation} \label{e^{(+)}}
e^{(+)}(q)=\frac{2}{L}\sum_{\tilde k}^{}
\frac{e^{-itk^2+ixk}-e^{-itq^2+ixq}}{k-q}.
\end{equation}
For the matrix $S^{(+)}$ all the calculation explained above on the
example of the matrix $S^{(-)}$ can be done similarly.
This leads to the structure (\ref{S+R+}) of the matrix $S^{(+)}$.
The matrix $R^{(+)}$ defined by (\ref{SR}) can be put into the form
(\ref{S+R+}) by means of the indentity (\ref{s=ee2}) applied to each
factor defining $R^{(+)}_{ab}$ in (\ref{SR}).

Let us now consider the thermodynamic properties of the functions
containing sums over quasimomenta $\tilde q$
(or $\tilde k$) in their definitions. These functions are $e^{(\pm)}$
entering the matrix elements of the matrices $S^{(\pm)}$, $R^{(+)}$ and
the function $g(x,t)$ entering the representations (\ref{NMV+}),
(\ref{NMV+2}).

Let us begin from the simplest example, the function $g$, which
is defined as
\begin{equation} %\label{}
g(x,t)=\frac{1}{L}\sum_{\tilde k}^{} e^{-itk^2+ixk},
\qquad k=\tilde k+\frac{\Lambda}{L}.
\end{equation}
In the limit $L\to \infty$ the values $(k)_j$ of the
quasimomenta $k$ fill the interval $(-\infty,+\infty)$ densely,
$(k)_{j+1}-(k)_j=(\tilde k)_{j+1}-(\tilde k)_j =2\pi/L$,
and the sum over values of $\tilde k$
should be changed for the integral,
\begin{equation} %\label{}
\frac{1}{L} \sum_{\tilde k}^{} f(\tilde k) :=
\frac{1}{L} \sum_{j=-\infty}^{\infty} f((\tilde k)_j) \to
\frac{1}{2\pi}\int\limits_{-\infty}^{\infty}dk\, f(k).
\end{equation}
Thus, in the limit $L\to\infty$ for the function $g$ one has
\begin{equation} %\label{}
g(x,t)\to G(x,t)=\frac{1}{2\pi} \int\limits_{-\infty}^{\infty}
dk\, e^{-itk^2+ixk}.
\end{equation}
The dependence on $\Lambda$ of the function $g$ disappears
in the limit $L\to \infty$. The same happens with respect to
the dependence on $m$ of functions $g_m:=g|_{\Lambda=\frac{2\pi
m}{N+1}}$.  All functions $g_m$ have the same thermodynamic limit which
is equal to function G.

Consider the function $e^{(+)}(q)$ given by (\ref{e^{(+)}}).
Recall that the sum is understood in the sense of prescription
described above. This prescription regularizes the sum on the
``infinities'' while the subtraction term, in the limit $L\to \infty$,
remove the singularity coming from the pole on the integration counter.
It means that such a sum turns into
the integral in the sense of the principal value, i.e., for the function
$e^{(+)}(q)$ one has
\begin{equation} %\label{}
e^{(+)}(q) \to E(q)
=P.v.\int\limits_{-\infty}^{\infty}dk\,
\frac{e^{-itk^2+ixk}}{\pi(k-q)}.
\end{equation}
Note that the dependence on $\Lambda$ and $\Theta$ of the function
$e^{(+)}(q)$ disappears in the limit $L\to\infty$ (function $E(q)$
does not depend on $\Lambda$ and $\Theta$) since $k$ and $q$
(function's argument) are assumed to run
the interval $(-\infty,+\infty)$ continuously.

The function $e^{(-)}(k)$ given by (\ref{e^{(-)}(k)}) has the same
thermodynamic limit as the function $e^{(+)}(k)$:
\begin{equation} %\label{}
e^{(-)}(k)\to E(k)
=P.v.
\int\limits_{-\infty}^{\infty}dq\,\frac{e^{-itq^2+ixq}}{\pi(q-k)}.
\end{equation}
Thus, all functions determining matrix elements of the matrices
$S^{(\pm)}$ and $R^{(+)}$ are well defined in the thermodynamic limit.

%%%%%%%%%%%%%%%%%%%%%%%%%%%%%%%%%%%%%%%%%%%%%%%%%%%%%%%%%%%%%%%%%%%%%%%


\begin{thebibliography}{99}
\bibitem{Tr1}
Tracy C.A., McCoy B.M.,
{\em Phys. Rev. Lett.\/} {\bf 31} (1973) 1500--1504.
\bibitem{Tr2}
Wu T.T., McCoyB.M., Tracy C.A., Barough E.,
{\em Phys. Rev.\/} {\bf B13} (1976) 316--374.
\bibitem{Tr3}
McCoy B.M., Tracy C.A., Wu T.T.,
{\em J. Math. Phys.} {\bf 18} (1977) 1058--1095.
\bibitem{Jimbo}
Jimbo A., Miwa T., Mori Y., Sato M.,
{\em Physica\/} {\bf D1} (1980) 80--158.
\bibitem{IIK-89}
Its A.R., Izergin A.G., Korepin V.E.,
{\em Phys. Lett.\/} {\bf 141} (1989) 121--125.
\bibitem{IIKS-90}
Its A.R., Izergin A.G., Korepin V.E., Slavnov N.A.,
{\em Int. J. Mod. Phys.\/} {\bf B4} (1990) 1003--1037.
\bibitem{IIKS-93}
Its A.R., Izergin A.G., Korepin V.E., Slavnov N.A.,
in: {\em Important developments in soliton theory\/}
(Springer Ser. Nonlinear Dynamics, Springer, 1993) 407--417.
\bibitem{KBI}
Korepin V.E., Bogoliubov N.M., Izergin A.G.,
{\em Quantum Inverse  Scattering Method And Correlation Functions\/}
(Cambridge University Press, 1993).
\bibitem{L-64}
Lenard A.,
{\em J. Math. Phys.\/} {\bf 5} (1964) 930--943.
\bibitem{L-66}
Lenard A.,
{\em J. Math. Phys.\/} {\bf 7} (1966) 1268--1272.
\bibitem{Yang}
Yang C.N.,
{\em Phys. Rev. Lett.\/} {\bf 19} (1967) 1312--1315.
\bibitem{Sutherland}
Sutherland B.,
{\em Phys. Rev.\/} {\bf B12} (1975) 3795--3805.
\bibitem{Gaudin}
Gaudin M.,
{\em La fonction d'Onde de Bethe\/}
(Masson, 1983).
\bibitem{Kulish}
Kulish P.P.,
{\em Physica\/} {\bf D3} (1981) 246--257.
\bibitem{Berkovich}
Berkorvich A.,
{\em J. Phys. A: Math. Gen.\/} {\bf 24} (1991) 1543--1556.
\bibitem{KS-90}
Korepin V.E., Slavnov N.A.,
{\em Commun. Math. Phys.\/} {\bf 129} (1990) 103--113.
\bibitem{KS-91}
Korepin V.E., Slavnov N.A.,
{\em Commun. Math. Phys.\/} {\bf 136} (1991) 633--646.
\bibitem{CIKT-92}
Colomo F., Izergin A.G., Korepin V.E., Tognetti V.,
{\em Phys. Lett.\/} {\bf A169} (1992) 243--247.
\bibitem{CIKT-93}
Colomo F., Izergin A.G., Korepin V.E., Tognetti V.,
{\em Theor. Math. Phys.\/} {\bf 94} N1 (1993) 11--38.
\bibitem{IP-97}
Izergin A.G., Pronko A.G.,
to be published in {\em Phys. Lett.\/} {\bf A}.
\end{thebibliography}
\end{document}